\documentclass[onecolumn,superscriptaddress,showpacs]{revtex4}

 \usepackage{graphicx}

\usepackage{bm,color}

\begin{document}

\title{Virial theorem for Onsager vortices in two-dimensional hydrodynamics}

\author{Pierre-Henri Chavanis}
\affiliation{ Laboratoire de Physique Th\'eorique (IRSAMC), CNRS and UPS,  Universit\'e de Toulouse, F-31062 Toulouse, France}


\begin{abstract}
We derive the virial theorem appropriate to two-dimensional
point vortices at statistical equilibrium in the microcanonical and
canonical ensembles.  In an unbounded domain, it relates the angular
velocity to the angular momentum and the temperature. Our expression
is valid for an arbitrary number of point vortices of possibly different
species. In the single-species case, and in the mean field
approximation, it reduces to the relation empirically obtained by
J.H. Williamson [J. Plasma Physics {\bf 17}, 85
(1977)].
\end{abstract}



\maketitle


\section{\label{intro}Introduction}

In a seminal paper, Onsager \cite{onsager} laid down the foundations
of the statistical mechanics of vortices in two-dimensional
hydrodynamics. He considered the point vortex gas as an idealization
of more realistic vorticity fields \footnote{The equations describing
the dynamics of point vortices are isomorphic to that of a 2D guiding
center plasma under a strong magnetic field \cite{jm,mj}. This analogy
has been the starting point of a number of analytical and numerical
studies in plasma physics.  In this paper, for the unity of the
presentation, we shall adopt the point vortex terminology even when we
refer to papers that have been developed in the plasma physics
context. The model of point vortices also describes the behavior of
vortex filaments with quantized circulation $h/m$ which appear in a
rotating container filled with superfluid helium.} and discovered
that negative temperature states are possible for this system. At
negative temperatures, corresponding to high energies, like-sign
vortices have the tendency to cluster into ``supervortices'' similar
to the large-scale vortices (e.g. Jupiter's great red spot) observed
in the atmosphere of giant planets.  The qualitative arguments of
Onsager were developed more quantitatively in a mean field
approximation by Joyce \& Montgomery
\cite{jm,mj}, Kida \cite{kida} and Pointin \& Lundgren \cite{pl,lp}, and by Onsager himself in
unpublished notes \cite{esree}. The statistical theory predicts that
the point vortex gas should relax towards an equilibrium state
described by the Boltzmann distribution.  Specifically, the
equilibrium stream function is solution of a Boltzmann-Poisson
equation. At positive temperatures, the Boltzmann-Poisson equation is
similar to the one appearing in the theory of electrolytes in plasma
physics (like-sign vortices ``repel'' each other) \cite{dh}. At
negative temperatures, it is similar to the one appearing in the
statistical mechanics of stellar systems (like-sign vortices
``attract'' each other) \cite{paddy,ijmpb}.  Many mathematical works
\cite{caglioti,k93,es,ca2,kl,sawada} have shown how a proper
thermodynamic limit could be rigorously defined for the point vortex
gas (in the Onsager picture). It is shown that the mean field
approximation becomes exact in the limit $N\rightarrow +\infty$ with
$\gamma\sim 1/N$, where $N$ is the number of point vortices and
$\gamma$ their individual circulation. The kinetic theory of the
point vortex gas has been developed in Refs. \cite{dubin,preR,pre,dubin2,cl,sano,bbgky,klim,kinonsager}.

The statistical equilibrium state of a single species system of point vortices in an unbounded domain was studied by Williamson \cite{williamson}. He used a mean field approximation and numerically solved the Boltzmann-Poisson equation for an axisymmetric distribution of point vortices. He empirically obtained a relation between the angular velocity, the angular momentum, the number of vortices and the temperature but did not manage to derive it. The same relation was obtained independently by Lundgren \& Pointin \cite{lp} who derived it from the microcanonical distribution of the point vortex gas by using essentially the same trick as the one used by Salzberg \& Prager \cite{spr,salzberg} to obtain the exact equation of state of a 2D plasma. On the other hand, Kiessling \cite{kiessling} derived the Williamson algebraic relation by a series of partial integrations for radial solutions of the Boltzmann-Poisson equation (he proved that the solutions are necessarily radially symmetric with decreasing density profile). He mentioned the similarity with a virial identity obtained by Corngold \cite{corngold} but also stressed that this identity is valid only at positive temperatures and corresponds to different types of averages. Therefore, its application to point vortices at negative temperatures is not clear.

In this paper, we provide a simple derivation of the virial theorem of
point vortices at statistical equilibrium and show that, for an
unbounded system, it is equivalent to the Williamson relation. Our
derivation is more general than previous ones since it is valid for a
multispecies gas of point vortices and goes beyond the mean field
approximation (furthermore, it does not assume, or use, the fact that
the flow is axisymmetric). Therefore, the relation that we obtain is
``exact''.  We also obtain a generalization of the virial theorem for
an axisymmetric system of point vortices enclosed within a disk in the
mean field approximation. However, in that case, the equation is not
closed (as it involves the pressure on the boundary). Finally, we
mention numerous analogies between 2D point vortices and 2D rotating
self-gravitating systems. In particular, the statistical equilibrium
state, the local equation of state, the condition of hydrostatic
equilibrium (in the rotating frame) and the virial theorem have a
similar form. The virial theorem plays a very important role in the
theory of simple liquids \cite{hansen} and in astrophysics
\cite{chandra}. To our knowledge, the ``virial of point vortices''
(especially at negative temperatures) has not been introduced before.
In view of its importance in other domains of physics, we think that
it is important to devote a specific paper to this
topic.

The paper is organized as follows. In Sec. \ref{pv}, we recall the
basic equations of the multi-species point vortex gas in an unbounded
domain. In Sec. \ref{vexact}, we consider the statistical equilibrium
state of point vortices in canonical and microcanonical ensembles, and
derive the Yvon-Born-Green (YBG) hierarchy and the ``exact'' virial
theorem. In Sec. \ref{sec_mfa}, we consider the mean field
approximation, and derive the multi-species Boltzmann-Poisson equation
(from the YBG hierarchy or from a maximum entropy principle) and the
mean field virial theorem.  We also consider some applications of the
virial theorem for point vortices in an unbounded domain or in a disc
(for axisymmetric distributions). The Appendices contain several
additional results. In Appendix \ref{vv}, we define and compute the
virial of point vortices. In Appendix \ref{sec_alt}, we provide
alternative derivations of the virial theorem in an unbounded domain
starting directly from the partition function or the density of
states, or from the explicit analytical solution of the
Boltzmann-Poisson equation (for a single species system). In Appendix
\ref{sec_nonrig}, we derive the virial theorem in a bounded domain for
purely logarithmic potentials (i.e. neglecting vortex images in the
case of point vortices), and comment on the meaning of the ``pressure
of point vortices''. In Appendix \ref{sec_yy}, we derive the two-body
correlation function and the virial theorem for a neutral spatially
homogeneous system of point vortices. Finally, in Appendix \ref{er},
we discuss the ``exact'' values of the critical inverse temperatures
beyond which point vortices form Dirac peaks.

\section{The two-dimensional point vortex gas}
\label{pv}

We consider a multi-species system of point vortices in two-dimensional hydrodynamics described by the Kirchhoff-Hamilton equations \cite{kirchhoff,newton}:
\begin{equation}
\label{pv1}
\gamma_i \frac{dx_i}{dt}=\frac{\partial H}{\partial y_i},\qquad \gamma_i \frac{dy_i}{dt}=-\frac{\partial H}{\partial x_i},
\end{equation}
where $\gamma_i$ is the circulation of point vortex $i$. In an
unbounded domain, the Hamiltonian can be written
\begin{equation}
\label{pv2}
H=\sum_{i<j} \gamma_{i} \gamma_{j} u(|{\bf r}_{i}-{\bf r}_{j}|),
\end{equation}
where
\begin{equation}
\label{pv3}
u(|{\bf r}-{\bf r}'|)=-\frac{1}{2\pi}\ln |{\bf r}-{\bf r}'|,
\end{equation}
is the potential of interaction between two point vortices which
satisfies the Poisson equation $\Delta u=-\delta$.  We note that the
coordinates $x$ and $y$ of the point vortices are canonically
conjugate. In a finite domain, this implies that their phase space is
bounded (it cannot exceed $V^{N}$ where $V$ is the volume (area) and
$N$ the number of point vortices). As first realized by Onsager
\cite{onsager}, this property leads to the occurrence of negative
temperature states at high energies. In an unbounded domain, the
effective size of the system is determined by the conservation of
angular momentum (see below), so that similar results are obtained.
An interesting feature of the dynamics (\ref{pv2})-(\ref{pv3}) is that
the interaction between point vortices determines the {\it velocity}
of a given vortex while the interaction between material particles
(like charges or stars) determines the {\it acceleration} (force by
unit of mass) of a given particle. This is linked to the absence of
kinetic energy (in the usual sense) in the Hamiltonian
(\ref{pv2}). This is another argument why negative temperatures are
possible for point vortices. For material particles, the temperature
is a measure of the average kinetic energy and it must be positive
\footnote{A related argument is as follows. Because of the kinetic
term $\sum_i \frac{1}{2}m_i v_i^2$ in the Hamiltonian of material
particles, the partition function $Z=\int e^{-\beta H}\, \prod_i d{\bf
r}_id{\bf v}_i$ can be normalizable only if the inverse temperature
$\beta=1/k_B T$ is positive. Since there is no kinetic part in the
Hamiltonian of point vortices, positive and negative temperatures are
possible.}.

For an isolated system, the energy $E=H$, the angular momentum $L=\sum_{i=1}^{N}\gamma_{i}{r}_{i}^2$ and the linear impulse ${\bf P}=-{\bf z}\times\sum_{i=1}^{N}\gamma_{i} {\bf r}$ are conserved (${\bf z}$ is a unit vector normal to the plane of the flow). Of course, the total number $N_a$ of point vortices of each species  is also conserved (the total circulation of species $a$ is $\Gamma_a=N_a\gamma_a$). The exact velocity of a point vortex located in ${\bf r}_{i}$ is given by
\begin{equation}
\label{pv4}
{\bf V}_{i}=-{\bf z}\times \frac{\partial\psi}{\partial {\bf r}_{i}}({\bf r}_i)=-{\bf z}\times\sum_{j\neq i} \gamma_{j}\frac{\partial u}{\partial {\bf r}_{i}}(|{\bf r}_{i}-{\bf r}_{j}|),
\end{equation}
where
\begin{equation}
\label{pv3b}
\psi({\bf r})=\sum_{i} \gamma_{i} u(|{\bf r}-{\bf r}_{i}|),
\end{equation}
is the stream function. The stream function is related to the
vorticity $\omega({\bf r})=\sum_i\gamma_i\delta({\bf r}-{\bf r}_i)$ by
the Poisson equation $\Delta\psi=-\omega$. Using Eq. (\ref{pv3}), the velocity
of a point vortex can be written explicitly
\begin{equation}
\label{pv5}
{\bf V}_{i}=\frac{1}{2\pi}{\bf z}\times
\sum_{j\neq i} \gamma_{j}\frac{{\bf r}_{i}-{\bf r}_{j}}{|{\bf r}_{i}-{\bf r}_{j}|^{2}}.
\end{equation}

We introduce the $N$-body distribution $P_N({\bf r}_1,...,{\bf r}_N,t)$ of the system giving the probability density of finding the first point vortex in ${\bf r}_1$, the second point vortex in ${\bf r}_2$... at time $t$. The normalization condition is $\int P_N \, d{\bf r}_1...d{\bf r}_N=1$. We introduce the energy, entropy, angular momentum and impulse functionals
\begin{eqnarray}
\label{pvnew1}
E[P_N]=\int P_N H\, \prod_{i}d{\bf r}_{i},\qquad  S[P_N]=-k_B\int P_N \ln P_N \, \prod_{i}d{\bf r}_{i},
\end{eqnarray}
\begin{eqnarray}
\label{pvnew2}
L[P_N]=\int P_N L\, \prod_{i}d{\bf r}_{i},\qquad {\bf P}[P_N]=\sum_{i=1}^{N}\int P_N {\bf P}\, \prod_{i}d{\bf r}_{i}.
\end{eqnarray}
We define the one- and two-body distribution functions by
\begin{equation}
\label{yc5}
P_{1}^{(a)}({\bf r}_{1},t)=\int P_{N}({\bf r}_{1},...,{\bf r}_{N},t)\, d{\bf r}_{2}...d{\bf r}_{N},
\end{equation}
\begin{equation}
\label{yc6}
P_{2}^{(ab)}({\bf r}_{1},{\bf r}_2,t)=\int P_{N}({\bf r}_{1},...,{\bf r}_{N},t)\, d{\bf r}_{3}...d{\bf r}_{N}.
\end{equation}
The average energy, angular momentum and impulse can then be written
\begin{eqnarray}
\label{pv6}
E=\frac{1}{2}\sum_{a,b} N_a(N_b-\delta_{ab}) \int P_2^{(ab)}({\bf r},{\bf r}',t) \gamma_{a} \gamma_{b} u(|{\bf r}-{\bf r}'|)\, d{\bf r}d{\bf r}',
\end{eqnarray}
\begin{equation}
\label{pv7}
{L}=\sum_{a}\int N_a P_1^{(a)}({\bf r},t) \gamma_{a} r^2\, d{\bf r},\qquad {\bf P}=-{\bf z}\times \sum_{a}\int N_a P_1^{(a)}({\bf r},t) \gamma_{a} {\bf r}\, d{\bf r}.
\end{equation}

{\it Remark 1:} We note that the angular momentum $L=\sum_i \gamma_i r_i^2$ is similar to the moment of inertia for material particles. On the other hand, the conservation of linear impulse ${\bf P}=-{\bf z}\times \sum_i \gamma_i {\bf r}_i$ is equivalent to the conservation of ${\bf R}=\sum_i \gamma_i {\bf r}_i$ which is similar to the center of mass for material particles. The conservation of ${\bf R}$ for point vortices implies that the ``center of vorticity'' is fixed while, for material particles, the center of mass has a rectilinear motion at constant velocity.

\section{The statistical equilibrium state}
\label{vexact}

We consider a system of point vortices at statistical equilibrium in an unbounded domain. In that case, the circulations of the point vortices must have the same sign otherwise they would form dipoles $(+,-)$ and ballistically escape to infinity so that no equilibrium state would be possible \cite{newton}.

\subsection{The YBG hierarchy in the canonical ensemble}
\label{yc}

We first consider the statistical equilibrium state of a point vortex
gas in the canonical ensemble (CE) in which the inverse temperature
$\beta=1/(k_B T)$ and the angular velocity $\Omega$ are prescribed (we
set the linear velocity ${\bf U}={\bf 0}$). For Hamiltonian systems
with long-range interactions, whose energy is non-additive, the
statistical ensembles may not be equivalent, even at the thermodynamic
limit $N\rightarrow +\infty$ \cite{cdr}.  Furthermore, for these
systems, the canonical ensemble has no physical sense because it is
not possible to define the notion of a thermal bath, especially at
negative temperatures \footnote{The canonical distribution represents
the statistical equilibrium state of a system of ``Brownian point
vortices'' \cite{bv} whose $N$-body dynamics is governed by stochastic
equations instead of the deterministic equations (\ref{pv1})-(\ref{pv2})
for usual Hamiltonian point vortices \cite{newton}. However, this
model of ``Brownian vortices'' is essentially academic.}.  However, it
is always possible to consider the canonical ensemble at a
mathematical level and study the corresponding equilibrium states (the
more relevant microcanonical ensemble will be considered in
Sec. \ref{ym}). This study is useful because, at the thermodynamic
limit, a canonical equilibrium state is always a microcanonical
equilibrium state with the corresponding value of the energy. The
converse is wrong in case of ensemble inequivalence.

The canonical $N$-body distribution is given by
\begin{equation}
\label{yc1}
P_{N}({\bf r}_{1},...,{\bf r}_{N})=\frac{1}{Z(\beta,\Omega)}e^{-\beta H-\beta \frac{\Omega}{2} {L}}.
\end{equation}
Because of the factor $e^{-\beta \frac{\Omega}{2} {L}}$, the distribution (\ref{yc1}) is  possibly normalizable only if
\begin{equation}
\label{yc3b}
{\rm sgn}(\gamma)\beta\Omega\ge 0.
\end{equation}
Using the normalization condition $\int P_N \, d{\bf r}_1...d{\bf r}_N=1$, the partition function is given by $Z(\beta,\Omega)=\int e^{-\beta H-\beta \frac{\Omega}{2} {L}} \, d{\bf r}_1...d{\bf r}_N$. The free energy is defined by $F(\beta,\Omega)=-\frac{1}{\beta}\ln Z(\beta,\Omega)$ and  the Massieu function by $J(\beta,\Omega)=-\beta F(\beta,\Omega)=\ln Z(\beta,\Omega)$.
From the canonical distribution (\ref{yc1}), it is easy to obtain the following expressions for the averages and the fluctuations of  energy and angular momentum:
\begin{equation}
\label{yc2}
\langle H\rangle =\frac{\partial (\beta F)}{\partial \beta},\quad \langle L\rangle =2\frac{\partial (\beta F)}{\partial (\beta\Omega)},
\end{equation}
\begin{equation}
\label{yc3}
\langle H^2\rangle -\langle H\rangle^2=k_B T^2 \frac{\partial \langle H\rangle}{\partial T},\quad \langle L^2\rangle -\langle L\rangle^2 = -2\frac{\partial \langle L\rangle}{\partial (\beta \Omega)},
\end{equation}
where the independent variables are $\beta$ and $\beta\Omega$. These expressions show in particular that the specific heat $C={\partial E}/{\partial T}$ is positive in the canonical ensemble (even at negative temperatures). Finally, we note that the canonical distribution (\ref{yc1}) minimizes the free energy functional $F[P_N]=E[P_N]-TS[P_N]+\frac{\Omega}{2}L[P_N]$ at fixed normalization. At statistical equilibrium, using Eq. (\ref{yc1}), we get $F[P_N^{eq}]=F(\beta,\Omega)$.

Differentiating Eq. (\ref{yc1}) with respect to ${\bf r}_{i}$, we obtain
\begin{eqnarray}
\label{yc4}
\frac{\partial P_{N}}{\partial {\bf r}_{i}}=-\beta P_{N}\left (\sum_{j\neq i}\gamma_{i}\gamma_{j}\frac{\partial u}{\partial {\bf r}_{i}}(|{\bf r}_i-{\bf r}_j|)+\Omega \gamma_{i}  {\bf r}_{i}\right ).
\end{eqnarray}
Taking $i=1$, integrating Eq. (\ref{yc4}) on $d{\bf r}_{2}...d{\bf r}_{N}$, and using Eqs. (\ref{yc5}) and (\ref{yc6}),  we obtain the first equation of the YBG hierarchy in the canonical ensemble
\begin{eqnarray}
\label{yc7}
\frac{\partial P_{1}^{(a)}}{\partial {\bf r}}({\bf r})=-\beta \sum_b (N_b-\delta_{ab}) \int   P_{2}^{(ab)}({\bf r},{\bf r}')\gamma_a\gamma_b\frac{\partial u}{\partial {\bf r}}(|{\bf r}-{\bf r}'|)\, d{\bf r}'
-\beta\Omega\gamma_{a}P_{1}^{(a)}({\bf r}) {\bf r}.
\end{eqnarray}

\subsection{The virial theorem from the canonical YBG hierarchy}
\label{vc}

We introduce the number density $n_a({\bf r})=N_a P_{1}^{(a)}({\bf r})$ of point vortices of species $a$  and the total number density $n({\bf r})=\langle \sum_i \delta({\bf r}-{\bf r}_i)\rangle=\sum_a n_a({\bf r})$. Similarly, we introduce the vorticity $\omega_a({\bf r})=N_a\gamma_{a}P_{1}^{(a)}({\bf r})=\gamma_a n_a({\bf r})$ of species $a$ and the total vorticity $\omega({\bf r})=\langle \sum_i \gamma_i\delta({\bf r}-{\bf r}_i)\rangle=\sum_a \omega_a({\bf r})$. It is also convenient to define the  local ``pressure'' \cite{houchesPH} by the relation
\begin{eqnarray}
\label{vc1}
p({\bf r})=n({\bf r}) k_{B}T=k_B T\sum_a \frac{\omega_a({\bf r})}{\gamma_a}.
\end{eqnarray}
This relation is similar to the local isothermal equation of state of a self-gravitating system in the canonical ensemble.   For point vortices, the pressure is positive at positive temperatures and negative at negative temperatures. Multiplying Eq. (\ref{yc7}) by $N_a$ and summing on the species, we obtain
\begin{eqnarray}
\label{vc2}
\nabla p({\bf r})=- \sum_{ab} N_a (N_b-\delta_{ab})\gamma_{a}\gamma_{b}\int P_{2}^{(ab)}({\bf r},{\bf r}') \frac{\partial u}{\partial {\bf r}}(|{\bf r}-{\bf r}'|)  \, d{\bf r}'-\Omega \omega({\bf r}) {\bf r}.
\end{eqnarray}
This identity is similar to  the condition of hydrostatic equilibrium for a self-gravitating system in the rotating frame (this analogy will become clearer in the mean field approximation, see Eq. (\ref{vmf1})). Taking the scalar product of Eq. (\ref{vc2}) with ${\bf r}$ and integrating over the whole domain, we obtain
\begin{eqnarray}
\label{vc3}
\int \nabla p\cdot {\bf r}\, d{\bf r}=-{\cal V}-\Omega L,
\end{eqnarray}
where ${\cal V}$ is the average value of the virial of the point vortex gas defined by Eq. (\ref{exv8}) and $L=\int\omega r^2\, d{\bf r}$ is the average angular momentum. Integrating the first term by parts with $PV\rightarrow 0$ at infinity, and using the isothermal equation of state (\ref{vc1}), we get
\begin{eqnarray}
\label{vc3b}
\frac{1}{2}\Omega L=Nk_B T-\frac{1}{2}{\cal V}.
\end{eqnarray}
Finally, using the expression (\ref{exv9}) of the virial in an unbounded domain, we obtain the exact virial theorem of point vortices at statistical equilibrium in the canonical ensemble
\begin{eqnarray}
\label{vc4}
\frac{1}{2}\Omega L=Nk_{B}(T-T_{c}),
\end{eqnarray}
with the exact critical temperature
\begin{eqnarray}
\label{vc5}
k_{B}T_{c}=-\frac{1}{8\pi N}\sum_{i=1}^N\sum_{j\neq i}\gamma_i\gamma_j=-\frac{1}{8\pi N}\left (\Gamma^2-\sum_{i=1}^N \gamma_i^2\right )=-\frac{1}{8\pi N}\left (\Gamma^2-\Gamma_2\right ),
\end{eqnarray}
where $\Gamma=\sum_a N_a\gamma_a$  is the total circulation and $\Gamma_2=\sum_a N_a\gamma_a^2$  the total ``enstrophy''. For a system of $N$ vortices with equal circulation $\gamma$, it reduces to
\begin{eqnarray}
\label{vc6}
k_{B}T_{c}=-(N-1)\frac{\gamma^{2}}{8\pi}.
\end{eqnarray}
We note that the critical temperature is negative. Equation (\ref{vc4}) can be derived directly from the partition function (see Appendix \ref{sec_az}). It generalizes the relation previously obtained by \cite{williamson,lp,kiessling} in the single species case and in the mean field approximation. Our approach clearly shows that this relation can be interpreted as ``the virial theorem of point vortices''. Some consequences of the virial theorem are discussed in Sec. \ref{sec_cons}.

\subsection{The YBG hierarchy in the microcanonical ensemble}
\label{ym}

We now consider the statistical equilibrium state of a system of point vortices in the microcanonical ensemble (MCE) where the energy $E$, the angular momentum $L$ and the center of vorticity ${\bf R}$ are fixed. The microcanonical ensemble  is the proper description of an isolated Hamiltonian system with long-range interactions \cite{cdr}.

The microcanonical $N$-body distribution is given by
\begin{eqnarray}
\label{ym1}
P_N({\bf r}_1,...,{\bf r}_N)=\frac{1}{g(E,{L})}\delta(E-H({\bf r}_1,...,{\bf r}_N))\delta\left ({L}-\sum_i \gamma_i r_i^2\right )\delta\left ({\bf R}-\sum_i \gamma_i {\bf r}_i\right ).
\end{eqnarray}
Using the normalization condition $\int P_N \, d{\bf r}_1...d{\bf r}_N=1$, the density of states is given by  $g(E,L)=\int \delta(E-H)\delta({L}-\sum_i \gamma_i r_i^2)\delta({\bf R}-\sum_i \gamma_i {\bf r}_i)\, d{\bf r}_{1}...d{\bf r}_{N}$. It does not depend on ${\bf R}$ since this constraint may be absorbed by a shift of origin in the integral \cite{lp}. We take the origin of the coordinates at the center of vorticity so that ${\bf R}={\bf 0}$. The entropy is defined by $S(E,L)=k_B \ln g(E,{L})$. Then, the temperature and the angular velocity are given by
\begin{eqnarray}
\label{ym2}
\frac{1}{T}=\frac{\partial S}{\partial E}, \qquad \frac{\Omega}{2T}=\frac{\partial S}{\partial {L}}.
\end{eqnarray}
Finally, we note that the microcanonical distribution (\ref{ym1}) maximizes the entropy functional $S[P_N]$ at fixed energy, angular momentum, impulse and normalization. At statistical equilibrium, using Eq. (\ref{ym1}), we get $S[P_N^{eq}]=S(E,L)$.

Differentiating Eq. (\ref{yc5}) with respect to ${\bf r}_{1}$ and using the microcanonical distribution (\ref{ym1}), we obtain
\begin{eqnarray}
\label{ym3}
\frac{\partial P_{1}^{(a)}}{\partial {\bf r}}({\bf r})=-\sum_{b} (N_b-\delta_{ab}) \int \frac{1}{g(E,{L})}\frac{\partial}{\partial E}\left (g(E,{L})P_2^{(ab)}({\bf r},{\bf r}')\right )\gamma_{a}\gamma_{b}\frac{\partial u}{\partial {\bf r}}(|{\bf r}-{\bf r}'|)\, d{\bf r}'\nonumber\\
-\frac{1}{g(E,{L})}\frac{\partial}{\partial {L}}\left (g(E,{L})P_1^{(a)}({\bf r})\right ) 2\gamma_{a}{\bf r}-\frac{1}{g(E,{L})}\frac{\partial}{\partial {\bf R}}\left (g(E,{L})P_1^{(a)}({\bf r})\right ) \gamma_{a}.
\end{eqnarray}
This is the first equation of the YBG hierarchy in the microcanonical ensemble \cite{lp}. This equation involves the quantities
\begin{eqnarray}
\label{ym4}
\frac{1}{g}\frac{\partial}{\partial E}\left (g P_2^{(ab)}\right )=\beta P_2^{(ab)}+\frac{\partial P_2^{(ab)}}{\partial E},
\end{eqnarray}
\begin{eqnarray}
\label{ym5}
\frac{1}{g}\frac{\partial}{\partial {L}}\left (g P_1^{(a)}\right )=\frac{1}{2}\beta \Omega P_1^{(a)}+\frac{\partial P_1^{(a)}}{\partial {L}},
\end{eqnarray}
\begin{eqnarray}
\label{ym6}
\frac{1}{g}\frac{\partial}{\partial {\bf R}}\left (g P_1^{(a)}\right )=\frac{\partial P_1^{(a)}}{\partial {\bf R}}.
\end{eqnarray}
The last terms in these expressions are difficult to evaluate. If we neglect these terms, Eq. (\ref{ym3}) reduces to
\begin{eqnarray}
\label{ym7}
\frac{\partial P_{1}^{(a)}}{\partial {\bf r}}({\bf r})=-\beta \sum_b (N_b-\delta_{ab}) \int   P_{2}^{(ab)}({\bf r},{\bf r}')\gamma_a\gamma_b\frac{\partial u}{\partial {\bf r}}(|{\bf r}-{\bf r}'|)\, d{\bf r}'
-\beta\Omega\gamma_{a}P_{1}^{(a)}({\bf r}) {\bf r},
\end{eqnarray}
and we recover the same equation as in the canonical ensemble [compare Eqs. (\ref{ym7}) and (\ref{yc7})]. However, further considerations show that the last terms in Eqs. (\ref{ym4})-(\ref{ym6}) can be neglected only in the $N\rightarrow +\infty$ limit (see Sec. \ref{sec_mfa}). Nevertheless, we show below that it is possible to obtain an exact explicit expression of the virial theorem in the microcanonical ensemble that is valid for any $N$.

\subsection{The virial theorem from the microcanonical YBG hierarchy}
\label{vmu}

Starting from the first equation (\ref{ym3}) of the YBG hierarchy and using a procedure similar to the one developed in Sec. \ref{vc}, we obtain
\begin{eqnarray}
\label{vmu1}
\frac{1}{k_B T}\int \nabla p\cdot {\bf r}\, d{\bf r}=-\frac{1}{g}\frac{\partial}{\partial E}(g{\cal V})-\frac{1}{g}\frac{\partial}{\partial L}(2gL)-\frac{1}{g}\frac{\partial}{\partial {\bf R}}(g{\bf R}),
\end{eqnarray}
where ${\cal V}$ is the average value of the virial of the point vortex gas defined by Eq. (\ref{exv8}), $L=\int\omega r^2\, d{\bf r}$ is the angular momentum, and ${\bf R}=\int\omega {\bf r}\, d{\bf r}$ is the center of vorticity. Expanding the derivatives and using Eq. (\ref{ym2}), the foregoing equation can be rewritten
\begin{eqnarray}
\label{vmu2}
\int \nabla p\cdot {\bf r}\, d{\bf r}=-{\cal V}-\Omega L-k_B T\frac{\partial {\cal V}}{\partial E}-4k_B T.
\end{eqnarray}
Integrating the first term by parts with $PV\rightarrow 0$ at infinity, and using the isothermal equation of state (\ref{vc1}), we get
\begin{eqnarray}
\label{vmu3}
\frac{1}{2}\Omega L=(N-2)k_B T-\frac{1}{2}{\cal V}-\frac{1}{2}k_B T\frac{\partial {\cal V}}{\partial E}.
\end{eqnarray}
Finally, using the expression (\ref{exv9}) of the virial in an unbounded domain, we obtain the exact virial theorem of point vortices at statistical equilibrium in the microcanonical ensemble
\begin{eqnarray}
\label{vmu4}
\frac{1}{2}\Omega L=Nk_B (T-T_c)-2k_B T,
\end{eqnarray}
where we have introduced the critical temperature (\ref{vc5}). The
same expression can be derived directly from the density of states
(see Appendix \ref{sec_vvg}). The result is slightly different than in
the canonical ensemble [see Eq. (\ref{vc4})] because of the term
$-2k_B T$. Furthermore, for a general potential of interaction $u$,
Eqs. (\ref{vc3b}) and (\ref{vmu3}) differ by an additional term
$-({1}/{2})k_B T {\partial {\cal V}}/{\partial E}$ that may not
vanish. However, the canonical and microcanonical expressions coincide
in the $N\rightarrow +\infty$ limit.

\section{The mean field approximation}
\label{sec_mfa}

\subsection{The Boltzmann-Poisson equation from the YBG hierarchy}
\label{sec_bpy}

We consider the thermodynamic limit  $N\rightarrow +\infty$ with $\gamma\sim 1/N$ \cite{caglioti,k93,es,ca2,kl,sawada}. This implies $E\sim 1$, $S\sim N$, $\beta\sim N$, $L\sim 1$ and $\Omega\sim 1$. Alternatively, one could introduce the dimensionless quantities $E_*=E/N^2\gamma^2$, $S_*=S/N$, $\beta_*=N\gamma^2\beta$, $L_*=L/N\gamma$ and $\Omega_*=\Omega/N\gamma$ that can be obtained from simple dimensional analysis (we note that the dimensional energy scales like $N^2$).  In this thermodynamic limit, it can be shown that the reduced correlation functions scale like $P_j'\sim (1/N)^{j-1}$ \cite{lp,bbgky}. In particular, the two-body correlation function scales like  $P_2'\sim 1/N$ (except in the region where $|{\bf r}-{\bf r}'|$ is small). Therefore, when $N\rightarrow +\infty$, the mean field approximation becomes exact:
\begin{eqnarray}
\label{bpy1}
P_{N}({\bf r}_1,...,{\bf r}_N)= \prod_{i=1}^N P_{1}({\bf r}_{i}).
\end{eqnarray}
In particular, we have
\begin{eqnarray}
\label{bpy2}
P_2^{(ab)}({\bf r},{\bf r}')=P_1^{(a)}({\bf r})P_1^{(b)}({\bf r}').
\end{eqnarray}
Furthermore, for $N\rightarrow +\infty$, the last terms in Eqs.  (\ref{ym4})-(\ref{ym6}) can be neglected, and we can make the approximation $N_a-1\simeq N_a$. Therefore, the first equation of the YBG hierarchy in microcanonical and canonical ensembles can be written
\begin{eqnarray}
\label{bpy3}
\nabla P_1^{(a)}({\bf r})=-\beta\gamma_a P_1^{(a)}({\bf r})\nabla\psi({\bf r})-\beta\gamma_a\Omega P_1^{(a)}({\bf r}) {\bf r},
\end{eqnarray}
where  $\psi({\bf r})=\left \langle \sum_i \gamma_i u(|{\bf r}-{\bf r}_i|)\right \rangle=\int \sum_b N_b \gamma_b P_1^{(b)}({\bf r}') u(|{\bf r}-{\bf r}'|)\, d{\bf r}'$ is the average stream function \cite{lp}. It is related to the average vorticity $\omega({\bf r})=\left \langle \sum_i \gamma_i \delta({\bf r}-{\bf r}_i)\right \rangle=\sum_b N_b \gamma_b P_1^{(b)}({\bf r})$ by
\begin{eqnarray}
\label{bpy3b}
\psi({\bf r})=\int u(|{\bf r}-{\bf r}'|)\omega({\bf r}')\, d{\bf r}',
\end{eqnarray}
which is the solution of the Poisson equation
\begin{eqnarray}
\label{bpy4}
\Delta\psi=-\omega,
\end{eqnarray}
with the Gauge condition $\psi+\frac{\Gamma}{2\pi}\ln r\rightarrow 0$ for $r\rightarrow +\infty$. The mean velocity of a
point vortex is ${\bf u}=-{\bf z}\times \nabla\psi$.

Equation (\ref{bpy3}) can be integrated to yield the Boltzmann distribution
\begin{eqnarray}
\label{bpy5}
\omega_a({\bf r})=N_a\gamma_a P_{1}^{(a)}({\bf r})=\Gamma_a\frac{ e^{-\beta\gamma_{a}\psi_{eff}({\bf r})}}{\int e^{-\beta\gamma_{a}\psi_{eff}({\bf r})}\, d{\bf r}},
\end{eqnarray}
where $\psi_{eff}({\bf r})=\psi({\bf r})+\frac{\Omega}{2}r^{2}$ is the relative stream function accounting for the conservation of angular momentum. The Boltzmann distribution (\ref{bpy5}) is  steady in a frame rotating with angular velocity $\Omega$. The vorticity profile decreases at large distances as
\begin{eqnarray}
\label{bpy5b}
\omega_a({\bf r})\propto e^{-\beta\gamma_a\frac{\Omega}{2}r^2}r^{\frac{\beta\gamma_a\Gamma}{2\pi}},\qquad (r\rightarrow +\infty).
\end{eqnarray}
Summing Eq. (\ref{bpy5}) on the species and substituting the resulting equation in Eq. (\ref{bpy4}), we obtain the multi-species Boltzmann-Poisson equation
\begin{eqnarray}
\label{bpy6}
-\Delta\psi=\sum_a \Gamma_a \frac{e^{-\beta\gamma_{a}\psi_{eff}({\bf r})}}{\int e^{-\beta\gamma_{a}\psi_{eff}({\bf r})}\, d{\bf r}}.
\end{eqnarray}
When $\beta>0$, the Boltzmann-Poisson equation (\ref{bpy6}) is similar to the one
appearing in the theory of electrolytes (like-sign vortices ``repel'' each other) \cite{dh} and when $\beta<0$, it is similar to the one
appearing in the statistical mechanics of stellar systems (like-sign vortices ``attract'' each other) \cite{paddy,ijmpb}. These analogies were noted in \cite{pre,esree}. We emphasize that these results are valid both in microcanonical and canonical ensembles.
Furthermore, Eqs. (\ref{bpy5}) and (\ref{bpy6}) remain valid  in a bounded domain ${\cal D}$, in which case the Boltzmann-Poisson equation must be solved with the boundary condition $\psi_{eff}=0$ on $\partial{\cal D}$ \cite{pl}.  In a disk, the angular momentum is conserved, as in an unbounded domain, and we get exactly the same equations  (with different boundary conditions).  If the domain does not possess the rotational invariance, the angular momentum is not conserved and $\Omega=0$. In a bounded domain, we can consider equilibrium states of point vortices with positive and negative circulations, since the boundary constrains the phase space to finite volumes.

In a neutral system of point vortices at positive temperatures, each
vortex has the tendency to be surrounded by vortices of opposite sign
which screen the interaction \cite{et,bv}. In that case, the system is
spatially homogeneous. In fact, if we linearize Eq. (\ref{bpy6}) for
$\beta\rightarrow 0$ (high temperatures), we obtain
$\Delta\psi-(\sum_a n_a\gamma_a^2)\beta\psi=0$ which shows that the
interaction is exponentially screened on a typical distance
$\lambda_D=\lbrack(\sum_a n_a\gamma_a^2)\beta\rbrack^{-1/2}$
corresponding to the Debye length in plasma physics. At
lower temperatures (higher $\beta$), the vortices have the tendency to
form pairs or dipoles  $(+,-)$, and the Debye-H\"uckel approximation is not
valid anymore (see Appendix \ref{er}). In a neutral system of point
vortices at negative temperatures, each vortex has the tendency to be
surrounded by vortices of the same sign. This corresponds to a form of
anti-shielding leading to the formation of clusters \cite{et,bv}.  For small $\beta$, we
obtain $\Delta\psi+(\sum_a n_a\gamma_a^2)|\beta|\psi=0$, so the effect
of negative temperatures is to convert the exponentially damped
linearized solutions discussed above into spatially oscillatory
ones \cite{mj}. At smaller $\beta$,
the system is spatially inhomogeneous and consists in
large-scale vortices (typically a big dipole). For a neutral
system with $N/2$ vortices of circulation $+\gamma$ and $N/2$ vortices
of circulation $-\gamma$, we have $\omega_+=A_+
e^{-\beta\gamma\psi_{eff}}$ and $\omega_-=A_-
e^{\beta\gamma\psi_{eff}}$. If we assume antisymmetry of the charge distribution about the center of the box, leading to $A_+=-A_-=A$, we obtain the celebrated
sinh-Poisson equation $\Delta\psi=2A\sinh(\beta\gamma\psi_{eff})$
\cite{mj,pl}. However, there is no fundamental reason why $A_+=-A_-$
should hold in general, so Eq. (\ref{bpy6}) is more general.

{\it Remark 2:} Kiessling \cite{kcrit} has shown that there exist an
inverse temperature $\beta_0<0$ above which the homogeneous solution
$\psi=0$ is the only solution of the Boltzmann-Poisson equation
(\ref{bpy6}) in the neutral case. In that case, the properties of the
system are adequately described by the two-body correlation function
(see Appendix \ref{sec_yy}). By contrast, for $\beta<\beta_0$, the
mean field equation (\ref{bpy6}) admits spatially inhomogeneous
solutions. Such solutions exist only above a critical inverse
temperature $\beta_c$ (see Appendix \ref{er}).

{\it Remark 3:} For point vortices in an unbounded domain, the
invariance by rotation of the system leads to a factor
$e^{-\frac{1}{2}\beta\gamma\Omega r^2}$ in the vorticity
distribution. When the condition (\ref{yc3b}) is satisfied, this
factor prevents the dispersion of the vortices and confines the
system. For rotating self-gravitating systems in an unbounded domain,
the equivalent factor in the density distribution is
$e^{\frac{1}{2}\beta m ({\bf \Omega}\times {\bf r})^2}$. This term,
which accounts for the effect of the centrifugal force, disperses the
particles. As a result, there is no statistical equilibrium state for
rotating self-gravitating systems in an unbounded domain, contrary to
the case of point vortices.

\subsection{The Boltzmann-Poisson equation from the maximum entropy principle}
\label{sec_bpe}

In the mean field approximation, the entropy, energy, angular momentum and circulation of each species can be written
\begin{eqnarray}
\label{bpe1}
S=-k_B\sum_{a}\int \frac{\omega_a}{\gamma_a}\ln \frac{\omega_a}{\gamma_a} \, d{\bf r},\quad E=\frac{1}{2}\int \omega\psi\, d{\bf r},\quad L=\int \omega r^2\, d{\bf r},\quad \Gamma_a=\int \omega_a \, d{\bf r}.
\end{eqnarray}
The entropy (\ref{bpe1}-a) can be obtained by substituting
Eq. (\ref{bpy1}) in Eq. (\ref{pvnew1}-b). It can also be obtained from
a standard combinatorial analysis \cite{jm,cl} starting from the
Boltzmann formula $S=k_B\ln W$, where $W$ is the number of {\it
microstates} (complexions) corresponding to a given {\it macrostate}
\footnote{A microstate is specified by the precise position $\lbrace
{\bf r}_1,...,{\bf r}_N\rbrace$ of the point vortices. A macrostate is
specified by the smooth vorticity field $\lbrace \omega_a({\bf
r})\rbrace$ giving the average number of point vortices of each
species in macrocells of size $\Delta$ satisfying $0<\Delta\ll
1$.}. Using the Stirling formula for $N\gg 1$, we obtain the
expression (\ref{bpe1}-a) of the Boltzmann entropy \cite{jm,cl}. The
energy (\ref{bpe1}-b) is obtained by substituting Eq. (\ref{bpy2}) in
Eq. (\ref{pv6}) and making the approximation $N_a-1\simeq N_a$.

In the microcanonical ensemble, the statistical equilibrium state is obtained by maximizing the Boltzmann entropy $S$ while conserving the energy $E$, the angular momentum $L$, and the total circulation $\Gamma_a$ of each species
\begin{eqnarray}
\label{nv3}
S(E,L,\Gamma_a)=\max_{\omega_a}\quad \lbrace S_{B}[{\omega_a}]\quad |\quad E[{\omega}]=E, \ L[{\omega}]=L, \ \Gamma_a[{\omega}_a]=\Gamma_a\rbrace.
\end{eqnarray}
This maximum entropy principle amounts to determining the {\it most probable state}, i.e. the one that is the most represented at the microscopic level. Introducing Lagrange multipliers to take the constraints into account, and writing the variational problem in the form $\delta S/k_B-\beta\delta E-\beta \frac{\Omega}{2}\delta {L}-\sum_a \alpha_a\delta \Gamma_a=0$, we find  \cite{jm,cl} that the critical points of constrained entropy are given by the mean field Boltzmann distribution  (\ref{bpy5}).

In the canonical ensemble, the statistical equilibrium state is obtained by maximizing the Boltzmann free energy $J=S/k_B-\beta E-\beta \frac{\Omega}{2}{L}$ (Massieu function) while conserving the total circulation $\Gamma_a$ of each species
\begin{eqnarray}
\label{nv3b}
J(\beta,\Omega,\Gamma_a)=\max_{\omega_a}\quad \lbrace J_{B}[{\omega_a}]\quad | \ \Gamma_a[{\omega}_a]=\Gamma_a \rbrace.
\end{eqnarray}
Writing the variational problem in the form $\delta J-\sum_a
\alpha_a\delta \Gamma_a=0$, we find that the critical points of
constrained free energy are also given by the mean field Boltzmann
distribution (\ref{bpy5}). Therefore, the series of equilibria
(critical points) are the same in MCE and CE but the nature of the
solutions (maxima, minima, saddle points) may differ in each
ensemble. When this happens, we speak of ensemble inequivalence
\cite{ellis}. It can be shown that a
solution of the canonical problem (\ref{nv3b}) is always a solution of
the more constrained dual microcanonical problem (\ref{nv3}) but the
reciprocal is wrong in case of ensemble inequivalence.

\subsection{The mean field virial theorem}
\label{sec_vmf}

Taking the logarithmic derivative of Eq. (\ref{bpy5}), we obtain
\begin{eqnarray}
\label{vv1}
\nabla\omega_a({\bf r})=-\beta \gamma_a\omega_a({\bf r})\nabla\psi_{eff}({\bf r}).
\end{eqnarray}
Dividing Eq. (\ref{vv1}) by $\gamma_a$, summing on the species and introducing the local pressure (\ref{vc1}),  we get
\begin{equation}
\label{vmf1}
\nabla p+\omega\nabla\psi_{eff}={\bf 0}.
\end{equation}
This equation is valid in microcanonical and canonical ensembles. It is similar to the condition of hydrostatic equilibrium for self-gravitating systems in a rotating frame \cite{bt}.  Taking the scalar product of Eq. (\ref{vmf1}) with ${\bf r}$, integrating over the entire domain and integrating by parts the pressure term, we obtain the mean field virial theorem
\begin{eqnarray}
\label{vmf2}
2\int p\, d{\bf r}-{\cal V}-\Omega L=2PV,
\end{eqnarray}
where ${\cal V}$ is the mean field virial (\ref{mfiv1}) of the point vortex gas, $V$ the area of the domain, and   $P=\frac{1}{2V}\oint p {\bf r}\cdot d{\bf S}$ the average pressure on the boundary of the domain (if the pressure is uniform on the boundary, with value $p_b$, then $P=p_b$). Using the isothermal equation of state (\ref{vc1}), we can rewrite the virial theorem as
\begin{eqnarray}
\label{vmf3}
2Nk_B T-{\cal V}-\Omega L=2PV.
\end{eqnarray}
In an unbounded domain, and for axisymmetric flows in a disk, the mean field virial ${\cal V}$ is given by Eq. (\ref{mfiv3}). In that case, the mean field virial theorem can be rewritten
\begin{eqnarray}
\label{vmf4}
PV=Nk_{B}(T-T_{c})-\frac{1}{2}\Omega L,
\end{eqnarray}
with the critical temperature
\begin{equation}
\label{vmf5}
k_{B}T_{c}=-\frac{\Gamma^{2}}{8\pi N}.
\end{equation}
For a single species system, the critical temperature reduces to
\begin{equation}
\label{vmf6}
k_{B}T_{c}=-\frac{N\gamma^{2}}{8\pi}.
\end{equation}
We see that the mean field critical temperature (\ref{vmf6}) only differs from the exact critical temperature  (\ref{vc6}) in CE by the replacement of $N-1$ by $N$. Of course, they coincide at the thermodynamic limit $N\rightarrow +\infty$ since the mean field approximation becomes exact in that limit.  For a neutral system ($\Gamma=0$), we find that
\begin{equation}
\label{vmf7}
k_{B}T_{c}=0.
\end{equation}
The relation (\ref{vmf4}) is valid in microcanonical and canonical ensembles. It is similar to the virial theorem of a rotating 2D self-gravitating system.  We now consider particular cases of the virial theorem.

(i) For an axisymmetric flow in a disk with $\Omega=0$, we obtain the equation of state
\begin{eqnarray}
\label{vmf8}
PV=Nk_{B}(T-T_{c}),
\end{eqnarray}
where $P=p(R)=n(R)k_B T$ and $V=\pi R^2$. For a single species system,
this relation can be directly derived from the analytical solution of
the 2D Boltzmann-Poisson system (see Appendix \ref{sec_abp}). On the
other hand, when $\Gamma=0$, Eq. (\ref{vmf8}) reduces to $PV=Nk_{B}T$
like for a perfect gas. This relation is of course trivial if the flow
is spatially homogeneous.

(ii) In an unbounded domain, $PV\rightarrow 0$ provided that the pressure decreases sufficiently rapidly with the distance. In that case, Eq. (\ref{vmf4}) reduces to
\begin{eqnarray}
\label{vmf9}
\frac{1}{2}\Omega L=Nk_{B}(T-T_{c}).
\end{eqnarray}
This returns the relation obtained in \cite{williamson,lp,kiessling} for the single species point vortex gas. However, our derivation is valid for a multi-species system of point vortices and does not assume (or use the fact) that the distribution is axisymmetric. The mean field virial theorem (\ref{vmf9}) has the same form as the exact virial theorem (\ref{vc4}) in CE and (\ref{vvg9}) in MCE, but the expression of the critical temperature is slightly different. However, all the expressions coincide at the thermodynamic limit $N\rightarrow +\infty$.

{\it Remark 4:} We note that the condition $L=0$ in MCE does not imply $\Omega=0$. There can be a rotation of the system even if the angular momentum is zero. Therefore, we must take the conservation of angular momentum into account  even when $L=0$.

\subsection{Summary}
\label{sec_sum}

Regrouping the previous results, we find that, in the mean field approximation, the equilibrium stream function is determined by the multi-species Boltzmann-Poisson equation (\ref{bpy6}) with appropriate boundary conditions. Then, the distribution of each species is determined by the Boltzmann distribution (\ref{bpy5}). These results are valid in MCE and CE. In CE, the temperature and the angular velocity are prescribed. In MCE, they must be expressed in terms of the energy and angular momentum that are the relevant control parameters (conserved quantities) in that case. This is done by substituting the Boltzmann distribution (\ref{bpy5}) in the constraints (\ref{bpe1}). In an unbounded domain,  we can use the virial theorem (\ref{vmf9}) to relate the angular velocity to the temperature according to
\begin{eqnarray}
\label{sum1}
\Omega=\frac{2N}{\beta L}\left (1-\frac{\beta}{\beta_c}\right ),
\end{eqnarray}
where $\beta_{c}=-8\pi N/\Gamma^{2}$. This equation replaces the constraint on the angular momentum (\ref{bpe1}-c). Then, the temperature is determined by the energy (\ref{bpe1}-b) and the chemical potentials by the circulation of each species (\ref{bpe1}-d). These equations generalize the equations given by Lundgren \& Pointin \cite{lp} for the single species point vortex gas in an unbounded domain.  The above procedure just determines critical points of entropy in MCE and critical points of free energy in CE. We must then study the second variations of these functionals to select {\it maxima} of entropy $S$ and free energy $J$. Minima or saddle points must be discarded. When several maxima exist for the same values of the constraints, we must distinguish stable (global) and metastable (local) states. Finally, phase transitions and possible situations of ensemble inequivalence must be considered.

\subsection{Consequences of the virial theorem in an unbounded domain}
\label{sec_cons}

We consider some consequences of the virial theorem (\ref{vmf9}) or (\ref{sum1}) in an unbounded domain. In that case, all the vortices must have the same sign. To be specific, we assume that their circulation is positive: ${\rm sgn}(\gamma)>0$.

\subsubsection{Microcanonical ensemble}
\label{sec_consm}

If $L=0$, the vorticity distribution is a Dirac peak at $r=0$ containing all the vortices: $\omega({\bf r})=\Gamma\delta({\bf r})$, and its energy is infinite $E\rightarrow +\infty$. This solution exists only at $\beta=\beta_c$ (i.e. $T=T_c$ ).  We now assume $L>0$ which is the generic case. Because of the inequality (\ref{yc3b}) and the virial theorem (\ref{sum1}), statistical equilibrium states can possibly exist only for $\beta\ge \beta_c$. If $\beta=\beta_c$, the angular velocity $\Omega=0$. In that case, the vorticity profile of species $a$ decreases as $r^{-4N\gamma_a/\Gamma}$ at large distances [see Eq. (\ref{bpy5b})]. This implies that the angular momentum is infinite \footnote{There is at least one population of point vortices for which $\gamma_a\le\Gamma/N$, implying an infinite value of $L_a$, then of $L$.} so this solution must be rejected. Therefore, statistical equilibrium states can possibly exist only for $\beta>\beta_c$ (i.e. $T\ge 0$ or $T<T_c$). In the single species case, Lundgren \& Pointin \cite{lp} have shown that the inverse temperature $\beta$ is a function of the augmented energy $\tilde E=8\pi E/\Gamma^2+\ln(L/\Gamma)$ only. Using the virial identity (\ref{sum1}), we conclude that $\Omega L$ is also a function of $\tilde E$ only. Let us consider particular limits.

(i) If $\beta\rightarrow \beta_c^+$ (i.e. $T\rightarrow T_c^{-}$), the angular velocity  $\Omega\sim -(\Gamma^2/4\pi L)(1-\beta/\beta_c)\rightarrow 0^{-}$. In the single species case, Lundgren \& Pointin \cite{lp} proposed an approximation of the vorticity profile of the form
\begin{eqnarray}
\label{consm1}
\omega(r)=\frac{\omega_0}{\left (1+\frac{\pi\omega_0}{\Gamma}\frac{\beta}{\beta_c}r^2\right )^2}e^{-\frac{\Gamma}{L}\left (1-\frac{\beta}{\beta_c}\right )r^2},
\end{eqnarray}
where $\omega_0$ is determined by the constraint $\Gamma=\int\omega\,
d{\bf r}$ leading to the relation $\pi L\omega_0/\Gamma^2 +\ln(\pi
L\omega_0/\Gamma^2)=-\gamma_E-\ln(1-\beta/\beta_c)$ where
$\gamma_E\simeq 0.557$ is the Euler constant. The first factor in
Eq. (\ref{consm1}) is the exact solution of the 2D Boltzmann-Poisson
equation with $\Omega=0$ (see Appendix \ref{sec_abp}) and the second
factor is a correction for large $r$ that reproduces the exact
asymptotic behavior (\ref{bpy5b}) of the density profile. In this
limit, the relation between the inverse temperature and the augmented
energy is $\beta/\beta_c=1-{\rm exp}\lbrack -1.5772-\tilde E-{\rm
exp}(1+\tilde E)\rbrack$. Therefore, $\beta$ tends towards $\beta_c$
very rapidly as $\tilde E\rightarrow +\infty$. For
$\beta\rightarrow \beta_c^+$, the system develops a Dirac peak
at $r=0$ containing almost all the vortices, surrounded by a
residual density profile with a Gaussian tail that ensures the
conservation of angular momentum.

(ii) If $\beta\rightarrow 0^{\pm}$ (i.e. $T\rightarrow \pm\infty$), the angular velocity $\Omega\sim 2N/\beta L\rightarrow \pm\infty$ and $\beta\Omega\rightarrow 2N/L$. In that case, the vorticity distribution of each species is Gaussian
\begin{eqnarray}
\label{consm2}
\omega_a(r)=\frac{N\gamma_a\Gamma_a}{\pi L}e^{-\frac{N\gamma_a}{L}r^2}.
\end{eqnarray}
In the single species case, the augmented energy corresponding to $\beta=0$ is $\tilde E(0)=\gamma_E-\ln 2\simeq -0.116$.

(iii) If $\beta\rightarrow +\infty$ (i.e. $T\rightarrow 0^+$), the angular velocity $\Omega\rightarrow \Gamma^2/(4\pi L)$. Since $\beta\psi_{eff}$ cannot diverge [see Eq. (\ref{bpy5})], we must have $\psi_{eff}\rightarrow 0$. This implies $\psi\sim -\frac{\Omega}{2}r^2$. In that case, the vorticity has a uniform value
\begin{eqnarray}
\label{consm3}
\omega=\frac{\Gamma^2}{2\pi L},
\end{eqnarray}
in a disk of radius $r_0=(2L/\Gamma)^{1/2}$, while it vanishes outside. The augmented energy corresponding to $\beta\rightarrow +\infty$ is $\tilde E(\beta=+\infty)=1/2-\ln 2\simeq -0.193$.

\begin{figure*}
\centering
\includegraphics[width=0.5\textwidth]{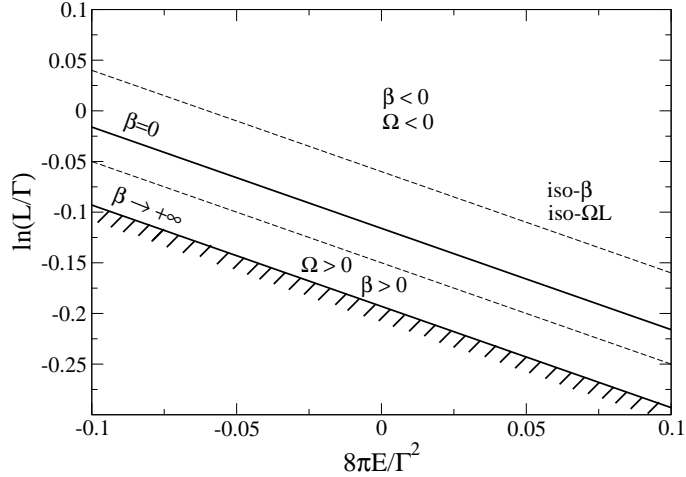}
\caption{Microcanonical parameter space $(E,L)$. The iso-$\beta$ and iso-$\Omega L$ lines correspond to $\tilde E={\rm Cst.}$ leading to the family of straight lines $\ln(L/\Gamma)=\tilde E-8\pi E/\Gamma^2$.}
\label{diagphasemicro}
\end{figure*}

Lundgren \& Pointin \cite{lp} have numerically solved the
Boltzmann-Poisson equation (\ref{bpy6}) for a single species point
vortex gas in an unbounded domain and they have computed all the
useful thermodynamic functions \footnote{They assumed that the vorticity field is axisymmetric. Actually, Kiessling  \cite{kiessling} has proven rigorously that only radial symmetric decreasing density profiles occur as solution of the 2D Boltzmann-Poisson equation in an unbounded domain.}. As a complement, we have represented the iso-$\beta$ lines and the iso-$\Omega L$ lines in the microcanonical parameter space $(E,L)$ in Fig. \ref{diagphasemicro}.

\subsubsection{Canonical ensemble}
\label{sec_consc}

Considering the inequality (\ref{yc3b})  and the virial theorem  (\ref{sum1}), we come to the following conclusions:

(i) If $\Omega=0$, statistical equilibrium states can possibly exist only at $T=T_{c}$ (i.e. $\beta=\beta_c$). The angular momentum is infinite while the energy can take arbitrary values. For a single species system, the vorticity profile is given in Appendix \ref{sec_abp}.

(ii) If $\Omega>0$, according to the virial theorem (\ref{sum1}), statistical equilibrium states can possibly exist only for $T\ge T_c$. In fact, according to the inequality (\ref{yc3b}), when $\Omega>0$ the canonical distribution can possibly  exist  only for $T\ge 0$ (i.e. $\beta\ge 0$) which is a more stringent condition than the one coming from the virial theorem.  For $\beta\rightarrow +\infty$, we must have $\psi_{eff}\rightarrow 0$ implying $\psi\sim -\frac{\Omega}{2}r^2$. In that case, the vorticity has a  uniform value $2\Omega$ in a disk of radius $r_0=(\Gamma/2\pi\Omega)^{1/2}$ while it vanishes outside. The angular momentum is $L=\Gamma^2/(4\pi\Omega)$ and the energy is $8\pi E/\Gamma^2=\tilde E(\beta=+\infty)-\ln (\Gamma/4\pi\Omega)$. For
$\beta\rightarrow 0^+$, we have $L\rightarrow +\infty$ and $E\rightarrow -\infty$, except if $\Omega\rightarrow +\infty$ such that $\beta\Omega$ is finite. In that case, the vorticity distribution of each species is Gaussian: $\omega_a=(\Gamma_a\gamma_a\beta\Omega/2\pi)e^{-\gamma_a\beta\Omega r^2/2}$. Furthermore, $L=2N/(\beta\Omega)$ and $8\pi E/\Gamma^2=\tilde E(0)-\ln(2/\gamma\beta\Omega)$ for a single species system.

(iii) If  $\Omega<0$, according to the virial theorem (\ref{sum1}), statistical equilibrium states can possibly exist only for $T\le T_c$ (i.e. $\beta_c\le \beta\le 0$) which is a more stringent condition than the inequality (\ref{yc3b}) yielding $T\le 0$ ($\beta\le 0$). If $T=T_c$ (i.e. $\beta=\beta_c$), the virial theorem implies $L=0$. In that case, the system forms a Dirac peak at $r=0$ containing all the vortices: $\omega({\bf r})=\Gamma \delta({\bf r})$, and the energy $E\rightarrow +\infty$. For
$\beta\rightarrow 0^-$, we have $L\rightarrow +\infty$ and $E\rightarrow -\infty$, except if $\Omega\rightarrow -\infty$ such that $\beta\Omega$ is finite. In that case, the vorticity distribution of each species is Gaussian: $\omega_a=(\Gamma_a\gamma_a\beta\Omega/2\pi)e^{-\gamma_a\beta\Omega r^2/2}$. Furthermore, $L=2N/(\beta\Omega)$ and $8\pi E/\Gamma^2=\tilde E(0)-\ln(2/\gamma\beta\Omega)$ for a single species system.

\begin{figure*}
\centering
\includegraphics[width=0.5\textwidth]{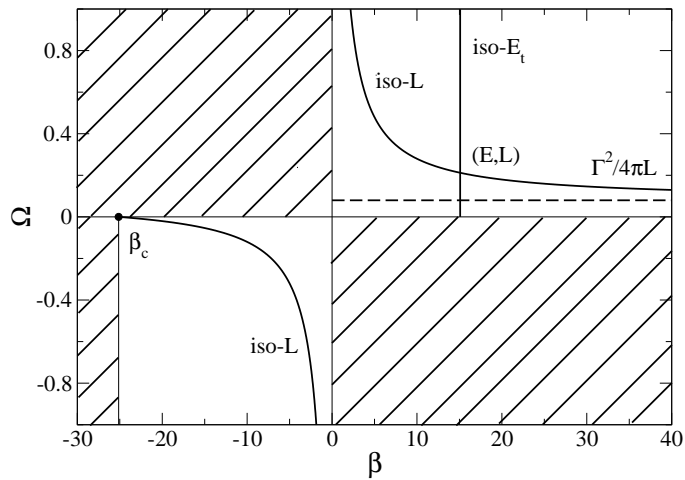}
\caption{Canonical parameter space $(\beta,\Omega)$. The iso-$L$ lines correspond to $\Omega=(2N/\beta L)(1-\beta/\beta_c)$, and they  tend to $\Gamma^2/(4\pi L)$ for $\beta\rightarrow +\infty$. The iso-$\tilde E$ lines correspond to $\beta={\rm Cst.}$ The intersection between these two curves determines the energy and the angular momentum $(E,L)$.}
\label{diagphasecano}
\end{figure*}

We have represented the iso-$\tilde E$ lines and the iso-$L$ lines in
the canonical parameter space $(\beta,\Omega)$ in
Fig. \ref{diagphasecano}. The ensembles are equivalent although the
relation between $(E,L)$ and $(\beta,\Omega)$ is not completely
trivial. In particular, there exist a critical point
$(\Omega=0,\beta=\beta_c)$ in the canonical ensemble corresponding to
an infinity of solutions with arbitrary energy ($-\infty<E<+\infty$)
and infinite angular momentum ($L=+\infty$) in the microcanonical
ensemble (see Appendix \ref{sec_abp}). This corresponds to a situation
of marginal ensemble equivalence \cite{ellis}.

\subsection{Consequences of the virial theorem for axisymmetric flows in a disk}
\label{sec_consb}

We consider some consequences of the virial theorem (\ref{vmf4}) for
an axisymmetric flow in a disk.

(i) If the vorticity field is a Dirac peak at $r=0$ containing all the point vortices, then $L=P=0$, implying $T=T_c$.

(ii) If $\Omega=0$, the virial theorem (\ref{vmf4}) reduces to Eq. (\ref{vmf8}). Since the pressure has the same sign as the temperature, an equilibrium state can possibly exist only for $\beta\ge \beta_c$ (i.e. $T\ge 0$ or $T\le T_c$). When $\Gamma=0$, implying $T_c=0$, the virial theorem does not put any constraint on $\beta$. If $P=0$ (e.g. if all the vortices form a Dirac peak at $r=0$), then $\beta=\beta_c$ (i.e. $T=T_c$). For the single species case, the vorticity profile is given in Appendix \ref{sec_abp}.

Smith \& O'Neil \cite{smith} have numerically solved the Boltzmann-Poisson equation (\ref{bpy6}) for a single species point vortex gas in a disk by taking the conservation of angular momentum into account. They discussed bifurcations between axisymmetric solutions and off-axis solutions, and reported situations of ensemble inequivalence. Caglioti {\it et al.} \cite{ca2}  proved under which conditions the equivalence of canonical and microcanonical ensembles holds.

{\it Remark 5:} In terms of the relative stream function $\psi_{eff}({\bf r})=\psi({\bf r})+\frac{\Omega}{2}r^2$, the Poisson equation can be written $\Delta\psi_{eff}=-\omega+2\Omega$. When $\Omega=\Omega_c\equiv \Gamma/(2\pi R^2)$, we have the analytical solution $\omega=\Gamma/(\pi R^2)$ and $\psi=-\Gamma r^2/(4\pi R^2)$, corresponding to $\psi_{eff}=0$. We can explicitly check that the virial theorem (\ref{vmf4}) is satisfied for this solution. In MCE, this solution exists only for $L=\Gamma R^2/2$ and $E=-\Gamma^2/(16\pi)$. In CE, it exists for $\Omega=\Omega_c$ at any temperature $\beta$.

\section{Conclusion}
\label{conc}

In this paper, we have shown that the ``Williamson relation'' \cite{williamson} between the angular velocity, the angular momentum and the temperature of the point vortex gas at statistical equilibrium in an unbounded domain can be interpreted as ``the virial theorem of point vortices''. Our approach  extends the virial identity obtained by Corngold \cite{corngold} which is only valid at positive temperatures and applies to a relatively different system. It also extends the calculations of Kiessling \cite{kiessling} that are valid for the single-species point vortex gas in the mean field approximation. We have obtained the ``exact'' expression of the virial theorem and of the critical temperature, valid beyond the mean field approximation, for a multi-species system of point vortices. We have discussed the differences between the canonical and microcanonical ensembles. We have also derived the mean field virial theorem of an axisymmetric distribution of point vortices in a disk although, in that case, the equation is not closed (as it involves the pressure on the boundary). Finally, we have discussed some consequences of the virial theorem  for the point vortex gas and derived  some necessary conditions of existence of statistical equilibrium.

\appendix

\section{The virial of point vortices}
\label{vv}

\subsection{The exact virial}
\label{exv}

We define the  exact virial of point vortices by
\begin{equation}
\label{exv1}
{\cal V}=\sum_{i=1}^{N}\gamma_{i} {\bf r}_{i}\cdot \nabla\psi({\bf r}_i).
\end{equation}
This quantity is similar to the virial of a self-gravitating system where the circulation plays the role of the mass and the stream function the role of the gravitational potential \cite{bt}. In an unbounded domain, using Eq.  (\ref{pv3b}), we obtain
\begin{equation}
\label{exv2}
{\cal V}=\sum_{i=1}^{N}\sum_{j\neq i}\gamma_{i}\gamma_{j} {\bf r}_{i} \cdot \frac{\partial u}{\partial {\bf r}_i}(|{\bf r}_i-{\bf r}_j|)=\sum_{i=1}^{N}\sum_{j\neq i}\gamma_{i}\gamma_{j} {\bf r}_{i} \cdot \frac{{\bf r}_{i}-{\bf r}_{j}}{|{\bf r}_{i}-{\bf r}_{j}|}u'(|{\bf r}_{i}-{\bf r}_{j}|),
\end{equation}
where we have used
\begin{equation}
\label{exv2bis}
\frac{\partial u}{\partial {\bf r}}(|{\bf r}-{\bf r}'|)=\frac{{\bf r}-{\bf r}'}{|{\bf r}-{\bf r}'|}u'(|{\bf r}-{\bf r}'|).
\end{equation}
Interchanging the dummy variables $i$ and $j$, and adding the
resulting expression to the previous one, the foregoing expression can be rewritten
\begin{equation}
\label{exv2tris}
{\cal V}=\frac{1}{2}\sum_{i=1}^{N}\sum_{j\neq i}\gamma_{i}\gamma_{j} |{\bf r}_{i}-{\bf r}_{j}| u'(|{\bf r}_{i}-{\bf r}_{j}|).
\end{equation}
Using the identity $u'(|{\bf x}|)=-1/(2\pi |{\bf x}|)$ coming from Eq. (\ref{pv3}), we obtain the simple exact result
\begin{eqnarray}
\label{exv4}
{\cal V}=-\frac{1}{4\pi}\sum_{i=1}^{N}\sum_{j\neq i} \gamma_{i}\gamma_{j},
\end{eqnarray}
which is independent on the configuration of the system. It only depends on the circulations of the point vortices. The sum can be rewritten
\begin{eqnarray}
\label{exv5}
\sum_{i=1}^{N}\sum_{j\neq i} \gamma_{i}\gamma_{j}=\sum_{i=1}^{N}\gamma_{i}(\Gamma-\gamma_{i})=\Gamma^{2}-\Gamma_2,
\end{eqnarray}
where $\Gamma=\sum_{i=1}^{N}\gamma_{i}$ is the total
circulation and $\Gamma_2=\sum_{i=1}^{N}\gamma_{i}^2$ is the total ``enstrophy''. For a system of $N$ identical point vortices with
circulation $\gamma$, we get
\begin{eqnarray}
\label{exv6}
{\cal V}=-N(N-1)\frac{\gamma^{2}}{4\pi}.
\end{eqnarray}
For a neutral system ($\Gamma=0$), we have ${\cal V}=\Gamma_2/(4\pi)$. In particular, for a neutral system with $N/2$ point vortices $+\gamma$ and  $N/2$ point vortices $-\gamma$, we obtain
\begin{eqnarray}
\label{exv7}
{\cal V}=\frac{N\gamma^{2}}{4\pi}.
\end{eqnarray}

The ensemble average value of the virial in an unbounded domain is given by
\begin{eqnarray}
\label{exv8}
{\cal V}=\sum_{ab} \gamma_{a}\gamma_{b}N_a (N_b-\delta_{ab})\int  P^{(ab)}_{2}({\bf r},{\bf r'})\, {\bf r}\cdot  \frac{\partial u}{\partial {\bf r}}(|{\bf r}-{\bf r}'|) \, d{\bf r}d{\bf r}'.
\end{eqnarray}
Using the same trick as before, the foregoing expression can be rewritten
\begin{eqnarray}
\label{exv8bis}
{\cal V}=\frac{1}{2}\sum_{ab} \gamma_{a}\gamma_{b}N_a (N_b-\delta_{ab})\int  P^{(ab)}_{2}({\bf r},{\bf r'})\, |{\bf r}-{\bf r}'| u'(|{\bf r}-{\bf r}'|) \, d{\bf r}d{\bf r}'.
\end{eqnarray}
Using the identity $u'(|{\bf x}|)=-1/(2\pi |{\bf x}|)$ coming from Eq. (\ref{pv3}), we obtain
\begin{eqnarray}
\label{exv9}
{\cal V}=-\frac{1}{4\pi}\sum_{ab} \gamma_{a}\gamma_{b}N_a(N_b-\delta_{ab})=-\frac{1}{4\pi}\left ( \Gamma^2-\Gamma_2\right),
\end{eqnarray}
where $\Gamma=\sum_a N_a\gamma_a$ and $\Gamma_2=\sum_a N_a\gamma_a^2$. Of course, this expression can be directly obtained from Eq. (\ref{exv4}).

The time average of the virial is
\begin{equation}
\label{exv1av1}
\langle {\cal V}\rangle =\lim_{\tau\rightarrow +\infty}\frac{1}{\tau}\int_0^{\tau}dt\, \sum_{i=1}^{N}\gamma_{i} {\bf r}_{i}(t)\cdot \nabla\psi({\bf r}_i(t)).
\end{equation}
Using the equations of motion (\ref{pv4}), it can be rewritten
\begin{equation}
\label{exv1av2}
\langle {\cal V}\rangle =\lim_{\tau\rightarrow +\infty}\frac{1}{\tau}\int_0^{\tau}dt\, \sum_{i=1}^{N}\gamma_{i} {\bf r}_{i}(t)\cdot ({\bf z}\times \dot{\bf r}_i(t)).
\end{equation}
Of course, the value of $\langle {\cal V}\rangle$ is given by Eq. (\ref{exv4}). However, the point that we want to make here is the following. If we integrate Eq. (\ref{exv1av2}) by parts, we get the trivial identity $\langle {\cal V}\rangle=\langle {\cal V}\rangle$. Therefore, this procedure does not give anything, contrary to the similar procedure used for material particles which leads to the usual virial theorem \cite{hansen}. In the present case, the virial theorem must be derived from the statistical equilibrium state. Indeed, for point vortices, the temperature is only defined at equilibrium while for material particles it can be defined at any time as the average kinetic energy \cite{hansen}. This implies that the virial theorem of material particles is valid for any steady state, while the virial theorem of point vortices is only valid at statistical equilibrium.

\subsection{The mean field virial}
\label{mfiv}

In the mean field approximation, the virial is
\begin{eqnarray}
\label{mfiv1}
{\cal V}=\int\omega({\bf r}) {\bf r}\cdot \nabla\psi({\bf r})\, d{\bf r}.
\end{eqnarray}
In an unbounded domain, using Eq. (\ref{bpy3b}), we obtain
\begin{eqnarray}
\label{mfiv2}
{\cal V}=\int\omega({\bf r})\omega({\bf r}') {\bf r}\cdot \frac{\partial u}{\partial {\bf r}}(|{\bf r}-{\bf r}'|)\, d{\bf r}d{\bf r}'=\int\omega({\bf r})\omega({\bf r}') {\bf r}\cdot \frac{{\bf r}-{\bf r}'}{|{\bf r}-{\bf r}'|}u'(|{\bf r}-{\bf r}'|)\, d{\bf r}d{\bf r}'. 
\end{eqnarray}
Interchanging the dummy variables ${\bf r}$ and ${\bf r}'$ and adding the
resulting expression to the previous one, the foregoing expression can be rewritten
\begin{eqnarray}
\label{mfiv2bis}
{\cal V}=\frac{1}{2}\int\omega({\bf r})\omega({\bf r}') |{\bf r}-{\bf r}'| u'(|{\bf r}-{\bf r}'|)\, d{\bf r}d{\bf r}'.
\end{eqnarray}
Using the identity $u'(|{\bf x}|)=-1/(2\pi |{\bf x}|)$ coming from Eq. (\ref{pv3}), we obtain
\begin{eqnarray}
\label{mfiv3}
{\cal V}=-\frac{\Gamma^{2}}{4\pi}.
\end{eqnarray}
In the single species case, it reduces to
\begin{eqnarray}
\label{mfiv4}
{\cal V}=-\frac{N^2\gamma^{2}}{4\pi}.
\end{eqnarray}
In the neutral case ($\Gamma=0$), we find
\begin{eqnarray}
\label{mfiv4b}
{\cal V}=0.
\end{eqnarray}

In a bounded domain, the virial has not a simple expression in general due to the presence of vortex ``images'' in the potential of interaction. There is, however, an exception. If we consider an axisymmetric vorticity profile in a disk, the Gauss theorem, obtained by integrating the Poisson equation (\ref{bpy4}), reads
\begin{eqnarray}
\label{mfiv5}
\frac{d\psi}{dr}=-\frac{\Gamma(r)}{2\pi r},
\end{eqnarray}
where $\Gamma(r)=\int_{0}^{r}\omega(r')2\pi r'\, dr'$ is the circulation within the disk of radius $r$. Then, the mean field virial (\ref{mfiv1}) is given by
\begin{eqnarray}
\label{mfiv6}
{\cal V}=-\frac{1}{2\pi}\int_0^R \Gamma(r)\frac{d\Gamma}{dr}\, dr=-\frac{1}{4\pi}\int_0^R \frac{d\Gamma^2}{dr}\, dr=-\frac{\Gamma^{2}}{4\pi},
\end{eqnarray}
like in an unbounded domain.

\section{Alternative derivations of the virial theorem}
\label{sec_alt}

\subsection{From the partition function}
\label{sec_az}

In the canonical ensemble, the exact virial theorem (\ref{vc4}) can be directly derived from the partition function $Z(\beta,\Omega)$ by making a simple transformation ${\bf r}\rightarrow {\bf r}/\sqrt{\Omega}$. This is similar to the procedure used by Salzberg \& Prager \cite{spr,salzberg} to derive the exact equation of state of a two-dimensional plasma. From the canonical distribution (\ref{yc1}), we find that the average value of the angular momentum can be written
\begin{equation}
\label{az1}
L =-\frac{2}{\beta}\frac{\partial \ln Z}{\partial\Omega},
\end{equation}
where
\begin{eqnarray}
\label{az2}
Z(\beta,\Omega)=\int e^{\frac{\beta}{2\pi}\sum_{i<j}\gamma_{i}\gamma_{j}\ln |{\bf r}_{i}-{\bf r}_{j}|}
 e^{-\beta\frac{\Omega}{2}\sum_{i}\gamma_{i}r_{i}^{2}}\, d{\bf r}_{1}...d{\bf r}_{N},
\end{eqnarray}
is the partition function. Making the change of variables ${\bf x}_{i}=\sqrt{\Omega} {\bf r}_{i}$ (we assume here $\Omega>0$; the case $\Omega<0$ can be treated similarly and leads to the same final result), the partition function can be rewritten
\begin{eqnarray}
\label{az3}
Z(\beta,\Omega)=\frac{1}{\Omega^{N}}e^{-\frac{\beta}{4\pi}\ln\Omega \sum_{i<j}\gamma_{i}\gamma_{j}} \int e^{\frac{\beta}{2\pi}\sum_{i<j}\gamma_{i}\gamma_{j}\ln |{\bf x}_{i}-{\bf x}_{j}|} e^{-\frac{\beta}{2}\sum_{i}\gamma_{i}x_{i}^{2}}\, d{\bf x}_{1}...d{\bf x}_{N}.
\end{eqnarray}
We note that the integral is now independent on $\Omega$. Therefore
\begin{eqnarray}
\label{az4}
\frac{\partial \ln Z}{\partial\Omega}=-\frac{\beta}{4\pi\Omega}\sum_{i<j}\gamma_{i}\gamma_{j}-\frac{N}{\Omega}.
\end{eqnarray}
Substituting this relation in Eq. (\ref{az1}) we obtain
\begin{eqnarray}
\label{az5}
\frac{1}{2}\Omega L=Nk_{B}(T-T_{c}),
\end{eqnarray}
with the critical temperature
\begin{eqnarray}
\label{az5b}
k_B T_c=-\frac{1}{8\pi N}\sum_{i=1}^N\sum_{j\neq i}\gamma_i\gamma_j.
\end{eqnarray}
These expressions coincide with Eqs. (\ref{vc4}) and (\ref{vc5}).

For an arbitrary potential of interaction $u$, the partition function is given by
\begin{eqnarray}
\label{lapf3new1}
Z(\beta,\Omega)=\int  e^{-\beta H-\beta\frac{\Omega}{2}L}\, d{\bf r}_{1}...d{\bf r}_{N}=\int  e^{-\beta\sum_{i<j}\gamma_{i}\gamma_{j} u(|{\bf r}_{i}-{\bf r}_{j}|)}e^{-\beta\frac{\Omega}{2}\sum_{i}\gamma_{i}r_{i}^{2}}\, d{\bf r}_{1}...d{\bf r}_{N}.
\end{eqnarray}
Making the change of variables  ${\bf x}_{i}=\sqrt{\Omega} {\bf r}_{i}$, we get
\begin{eqnarray}
\label{lapf3new2}
Z(\beta,\Omega)=\frac{1}{\Omega^N}\int  e^{-\beta\sum_{i<j}\gamma_{i}\gamma_{j} u(\frac{1}{\sqrt{\Omega}}|{\bf x}_{i}-{\bf x}_{j}|)}e^{-\frac{\beta}{2}\sum_{i}\gamma_{i}x_{i}^{2}}\, d{\bf x}_{1}...d{\bf x}_{N}.
\end{eqnarray}
Taking the derivative of this expression with respect to $\Omega$, using Eq. (\ref{az1}), and restoring the original coordinates, we obtain
\begin{eqnarray}
\label{lapf3new3}
\frac{1}{2}\Omega L=Nk_B T-\frac{1}{2} \int \, d{\bf r}_{1}...d{\bf r}_{N} \frac{1}{Z(\beta,V)} e^{-\beta H-\beta\frac{\Omega}{2}L} \sum_{i<j}\gamma_{i}\gamma_{j} |{\bf r}_{i}-{\bf r}_{j}| u'(|{\bf r}_{i}-{\bf r}_{j}|).
\end{eqnarray}
In the last term, we recognize the canonical average of the virial (\ref{exv2tris}). Therefore, Eq. (\ref{lapf3new3}) can be rewritten
\begin{eqnarray}
\label{lapf3new4}
\frac{1}{2}\Omega L=Nk_B T-\frac{1}{2}{\cal V}.
\end{eqnarray}
This coincides with the virial theorem (\ref{vc3b}) obtained from the canonical YBG equation. 
For the potential of interaction (\ref{pv3}), the virial ${\cal V}$ is given by Eq.  (\ref{exv4}), leading to Eq. (\ref{az5}).

\subsection{From the density of states}
\label{sec_vvg}

In the microcanonical ensemble, the virial theorem can be directly derived from the density of states $g(E,L)$ by making a simple transformation ${\bf r}\rightarrow \sqrt{L}{\bf r}$ \cite{lp}. Taking the origin at the center of vorticity, so that ${\bf R}={\bf 0}$, the density of states can be written
\begin{eqnarray}
\label{vvg2}
g(E,L)=\int \delta \left (E+\frac{1}{2\pi}\sum_{i<j}\gamma_{i}\gamma_{j}\ln |{\bf r}_{i}-{\bf r}_{j}|\right )\delta \left (L-\sum_{i=1}^{N} \gamma_{i}r_{i}^{2}\right )\delta \left (\sum_{i=1}^{N}\gamma_{i}{\bf r}_{i}\right )\, d{\bf r}_{1}...d{\bf r}_{N}.
\end{eqnarray}
Making the change of variables ${\bf x}_{i}={\bf r}_{i}/\sqrt{L}$, we obtain
\begin{eqnarray}
\label{vvg3}
g(E,L)=L^{N-2}\int \delta \left (E'+\frac{1}{2\pi}\sum_{i<j}\gamma_{i}\gamma_{j}\ln |{\bf x}_{i}-{\bf x}_{j}|\right )\delta \left (1-\sum_{i=1}^{N}\gamma_{i}x_{i}^{2}\right)\delta \left (\sum_{i=1}^{N}\gamma_{i}{\bf x}_{i}\right )\, d{\bf x}_{1}...d{\bf x}_{N},
\end{eqnarray}
with
\begin{eqnarray}
\label{vvg4}
E'=E+\frac{1}{4\pi}\ln L \sum_{i<j}\gamma_{i}\gamma_{j},
\end{eqnarray}
where we have used the identity $\delta(\lambda
x)=\frac{1}{|\lambda|}\delta(x)$. Therefore, the density of states is
of the form $g(E,L)=L^{N-2}g(E',1)$ and the entropy $S(E,L)=k_{B}\ln
g(E,L)$ satisfies the relation
\begin{eqnarray}
\label{vvg6}
S(E,L)=(N-2)k_{B}\ln L+S(E',1).
\end{eqnarray}
Accordingly,
\begin{eqnarray}
\label{vvg7}
\frac{\partial S}{\partial L}=(N-2)k_{B}\frac{1}{L}+\frac{\partial S}{\partial E}(E',1)\frac{1}{4\pi L}\sum_{i<j}\gamma_{i}\gamma_{j},
\end{eqnarray}
and
\begin{eqnarray}
\label{vvg8}
\frac{\partial S}{\partial E}=\frac{\partial S}{\partial E}(E',1).
\end{eqnarray}
Using the relations (\ref{ym2}), we finally obtain the microcanonical virial theorem
\begin{eqnarray}
\label{vvg9}
\frac{1}{2}L\Omega=(N-2)k_{B}T+\frac{1}{8\pi}\sum_{i=1}^{N}\sum_{j\neq i}\gamma_{i}
\gamma_{j}.
\end{eqnarray}
It slightly differs from the canonical virial theorem   (\ref{az5})-(\ref{az5b}) due to the factor $(N-2)$ instead of $N$ in front of $k_B T$. Our expression (\ref{vvg9}) also slightly differs from the one obtained by Lundgren \& Pointin \cite{lp} which involves $N-1$ instead of $N-2$ (apparently, they forgot the contribution arising from the linear impulse). Of course, in the thermodynamic limit $N\rightarrow +\infty$, all the expressions agree.

For an arbitrary potential of interaction $u$, the density of states is given by
\begin{eqnarray}
\label{lapf3new5}
g(E,L)=\int \delta \left (E-\sum_{i<j}\gamma_{i}\gamma_{j} u(|{\bf r}_{i}-{\bf r}_{j}|)\right )\delta \left (L-\sum_{i=1}^{N} \gamma_{i}r_{i}^{2}\right )\delta \left (\sum_{i=1}^{N}\gamma_{i}{\bf r}_{i}\right )\, d{\bf r}_{1}...d{\bf r}_{N}.
\end{eqnarray}
Making the change of variables ${\bf x}_{i}={\bf r}_{i}/\sqrt{L}$, we get 
\begin{eqnarray}
\label{lapf3new6}
g(E,L)=L^{N-2}\int \delta \left (E-\sum_{i<j}\gamma_{i}\gamma_{j} u(\sqrt{L}|{\bf x}_{i}-{\bf x}_{j}|)\right )\delta \left (1-\sum_{i=1}^{N}\gamma_{i}x_{i}^{2}\right)\delta \left (\sum_{i=1}^{N}\gamma_{i}{\bf x}_{i}\right )\, d{\bf x}_{1}...d{\bf x}_{N}.
\end{eqnarray}
Taking the derivative of this expression with respect to $L$, using Eq. (\ref{ym2}), and restoring the original coordinates, we obtain
\begin{eqnarray}
\label{lapf3new7}
\frac{L\Omega}{2k_B T}=N-2-\frac{1}{2}\frac{1}{g(E,L)}\frac{\partial}{\partial E} g(E,L)\int \, d{\bf r}_{1}...d{\bf r}_{N} \frac{1}{g(E,L)}\delta \left (E-\sum_{i<j}\gamma_{i}\gamma_{j} u(|{\bf r}_{i}-{\bf r}_{j}|)\right )\nonumber\\
\times\delta \left (L-\sum_{i=1}^{N} \gamma_{i}r_{i}^{2}\right )\delta \left (\sum_{i=1}^{N}\gamma_{i}{\bf r}_{i}\right ) \sum_{i<j}\gamma_{i}\gamma_{j} |{\bf r}_{i}-{\bf r}_{j}| u'(|{\bf r}_{i}-{\bf r}_{j}|).
\end{eqnarray}
In the last term, we recognize the microcanonical average of the virial (\ref{exv2tris}). Therefore, Eq. (\ref{lapf3new7}) can be rewritten
\begin{eqnarray}
\label{lapf3new8}
\frac{L\Omega}{2k_B T}=N-2-\frac{1}{2}\frac{1}{g(E,L)}\frac{\partial}{\partial E} (g(E,L){\cal V}).
\end{eqnarray}
Finally, expanding the derivative, and using Eq. (\ref{ym2}), we obtain
\begin{eqnarray}
\label{lapf3new9}
\frac{1}{2}L\Omega=(N-2)k_B T-\frac{1}{2}{\cal V}-\frac{1}{2}k_B T\frac{\partial {\cal V}}{\partial E}.
\end{eqnarray}
This coincides with the virial theorem (\ref{vmu3}) obtained from the microcanonical YBG equation. For the potential of interaction (\ref{pv3}), the virial ${\cal V}$ is given by Eq.  (\ref{exv4}), leading to Eq. (\ref{vvg9}).

\subsection{From the 2D Boltzmann-Poisson equation}
\label{sec_abp}

For a single-species point vortex gas with $\Omega=0$, the
Boltzmann-Poisson equation (\ref{bpy6}) can be solved
analytically in the axisymmetric case \cite{lp,williamson,kiessling,caglioti,houchesPH}. In addition to 2D hydrodynamics, this analytical profile appeared in various domains such as plasma physics \cite{bennett,kl2}, astrophysics \cite{stodolkiewicz,ostriker,klb,paddy,ap,sc} and biology \cite{cp,hv,bkln,bio}. Actually, the 2D Boltzmann-Poisson equation is also known as Liouville's equation  in differential geometry and the analytical axisymmetric profile presented below appears as a special case in a general formula of Liouville \cite{liouville}.

(i) In an unbounded domain, the solutions exist only at the temperature
$k_B T_c=-\Gamma\gamma/8\pi$ and they are given by (see, e.g., \cite{bio}):
\begin{eqnarray}
\label{av1}
\omega(r)=\frac{\omega_0}{\left (1+\frac{\pi\omega_0}{\Gamma}r^2\right )^2},\qquad \psi(r)=-\frac{\Gamma}{4\pi}\ln\left (\frac{\Gamma}{\pi\omega_{0}}+r^2\right ),
\end{eqnarray}
where $\omega_0$ is the central density and we have used the Gauge condition $\psi(r)+\frac{\Gamma}{2\pi}\ln r\rightarrow 0$ for $r\rightarrow +\infty$. When $\omega_0\rightarrow 0$, the distribution is almost uniform and when $\omega_0\rightarrow +\infty$, it tends to a Dirac peak: $\omega({\bf r})\rightarrow \Gamma\delta({\bf r})$.  Since $\omega\sim r^{-4}$
for $r\rightarrow +\infty$, the angular momentum is infinite (except
when $\omega({\bf r})=\Gamma\delta({\bf r})$ in which case $L=0$). The energy is given by
\begin{eqnarray}
\label{av2}
E=-\frac{\Gamma^2}{8\pi}\left\lbrack 1+\ln\left (\frac{\Gamma}{\pi\omega_0}\right )\right\rbrack,
\end{eqnarray}
and it goes from $-\infty$ (as $\omega_0\rightarrow 0$) to $+\infty$ (as $\omega_0\rightarrow +\infty$). The caloric curve is a straight
line $T(E)=T_c$.  The entropy is given as a function of the energy by the linear relation
\begin{eqnarray}
\label{av3}
S=k_{B}\left \lbrace N-\frac{8\pi E}{\Gamma\gamma}+N\ln\left (\frac{\pi}{N}\right )\right \rbrace.
\end{eqnarray}
The free energy $J=S-\beta E$ is given by
\begin{eqnarray}
\label{av4}
J=N\left\lbrack 1+\ln\left (\frac{\pi}{N}\right )\right\rbrack,
\end{eqnarray}
and it has the same value for all the solutions, whatever the value of
$\omega_0$.  In MCE, there is only one solution for each value of the
energy $E$, and the set of all these solutions has the same
temperature $T=T_c$. These solutions are stable (global entropy maxima
at fixed energy and circulation). In CE, the system exhibits an
infinity of solutions at $T=T_c$ that are parameterized by the central
vorticity.  There is no solution for other values of the
temperature. These solutions are metastable. Therefore, if we relax
the constraint on the angular momentum (this amounts to authorizing
$L=\infty$), implying $\Omega=0$, there is non-uniqueness in the
canonical ensemble while there is uniqueness in the microcanonical
ensemble. This is a situation of {\it marginal ensemble equivalence}
\cite{ellis}. The entropy $S(E)$ is a straight line, so it is both
concave and convex. The inverse specific heat
$d^2S/dE^2=d\beta/dE$ vanishes. The line of microcanonical
solutions $E\in \rbrack -\infty, +\infty\lbrack$ is mapped to a single
point $\beta=\beta_c$ in the canonical ensemble. For a recent example
of marginal ensemble inequivalence in the context of fluid mechanics, see
\cite{herbert}.

(ii) In a bounded domain, the solutions exist only for
$T\ge 0$ or $T\le T_c=-\Gamma\gamma/8\pi$ (i.e. $\beta\ge\beta_c=-8\pi/\Gamma\gamma$) and they are given by (see, e.g., \cite{bio}):
\begin{eqnarray}
\label{av5}
\omega(r)=\frac{1}{\pi R^2}\frac{\Gamma}{1-T_c/T}\frac{1}{\left\lbrack 1+\frac{T_c}{T-T_c}\left (\frac{r}{R}\right )^2\right \rbrack^2},
\end{eqnarray}
\begin{eqnarray}
\label{av6}
\psi(r)=\frac{2k_{B}T}{\gamma}\ln\left\lbrack 1-\frac{T_c}{T}+\frac{T_c}{T}\left (\frac{r}{R}\right )^2\right\rbrack,
\end{eqnarray}
where we have used the Gauge condition $\psi(R)=0$. From
Eq. (\ref{av5}), writing $P=p(R)=\omega(R)k_B T/\gamma$ and $V=\pi
R^2$, we obtain
\begin{eqnarray}
PV=Nk_B(T-T_c),
\end{eqnarray}
and we recover the equation of state (\ref{vmf8}). The central
vorticity is given by $\omega_0=(\Gamma/\pi R^2)(1-T_c/T)^{-1}$ and
the angular momentum by $L=\Gamma R^2(T/T_c-1)\lbrack
-(T/T_c)\ln(1-T_c/T)-1\rbrack$.  When $T>0$, the vorticity increases
with the distance and the vortices tend to accumulate at the boundary
(the interaction is ``repulsive'' like between like-sign charges in
electrolytes). When $T<0$, the vorticity decreases with the distance
and the vortices tend to cluster at the center of the domain (the
interaction is ``attractive'' like between stars in galaxies). When
$T\rightarrow \pm\infty$ (i.e. $\beta=0$), the distribution is
uniform. When $T=0$ (i.e. $\beta\rightarrow +\infty$), the
distribution is a Dirac peak at $r=R$.  When $T=T_c$
(i.e. $\beta=\beta_c$), the distribution is a Dirac peak at
$r=0$. Some vorticity profiles are represented in Figs. 4 and 5 of
\cite{houchesPH}. The energy is given by
\begin{eqnarray}
\label{av7}
E=-\frac{\Gamma^2}{8\pi}\left\lbrack \frac{T}{T_c}+\left (\frac{T}{T_c}\right )^2\ln\left (\frac{T-T_c}{T}\right )\right\rbrack,
\end{eqnarray}
and it goes from $0$ (for $T=0$) to $+\infty$ (as $T\rightarrow
T_c$). The transition between positive and negative temperatures
(corresponding to $T\rightarrow \pm\infty$, i.e. $\beta=0$) takes place at the energy
$E=\Gamma^2/(16\pi)$.   The entropy is given by
\begin{eqnarray}
\label{av8}
S=k_{B}\left \lbrace -N\ln\left (\frac{N}{V}\right )+2N+N\left (2\frac{T}{T_c}-1\right )\ln \left (1-\frac{T_c}{T}\right )\right \rbrace.
\end{eqnarray}
Eliminating $T$ between Eqs. (\ref{av7}) and (\ref{av8}), we get $S(E)$. The caloric curve $T(E)$ and the entropy curve $S(E)$ are plotted in Figs. 6 and 7 of \cite{houchesPH}. Finally, the free energy $J=S-\beta E$ is given as a function of the temperature by
\begin{eqnarray}
\label{av9}
J=k_B\left \lbrace -N \ln\left (\frac{N}{V}\right )+N+N\left (\frac{T}{T_c}-1\right )\ln\left (1-\frac{T_c}{T}\right )\right \rbrace.
\end{eqnarray}
In MCE, the solutions exist for any $E$ and they have an inverse
temperature $\beta\ge \beta_c$. In CE, the solutions exist for any
$\beta\ge \beta_c$. When the angular momentum constraint is relaxed
($\Omega=0$), the ensembles are equivalent.

\section{Virial theorem in a bounded domain for purely logarithmic interactions}
\label{sec_nonrig}

We consider a system of point vortices in a bounded domain of area $V$. We give here a non-rigorous derivation of the virial theorem by ignoring  the contribution of ``vortex images'' due to boundaries. In other words, we use the potential of interaction (\ref{pv3}) as in an unbounded domain. This approximation has also been made by  Edwards \& Taylor \cite{et}. It is difficult to say to which extent this approximation is accurate. However, one motivation of the following calculations is to help interpreting the curious notion of ``pressure of point vortices'' that we have introduced heuristically in Sec. \ref{vc}. On the other hand, the following results are exact for particles interacting by a purely logarithmic potential $u(|{\bf r}-{\bf r}'|)=-\frac{1}{2\pi}\ln|{\bf r}-{\bf r}'|$ in 2D, like plasmas and self-gravitating systems.

\subsection{From the YBG hierarchy}
\label{nrybg}

If we ignore the contribution of vortex images, the first equation of the canonical YBG hierarchy is given by Eq. (\ref{yc7}). If we define the local pressure by Eq. (\ref{vc1}), we obtain Eq. (\ref{vc3}). Integrating the first term by parts, introducing the average pressure
\begin{eqnarray}
\label{nr1}
P=\frac{1}{2V}\oint p{\bf r}\cdot d{\bf S},
\end{eqnarray}
on the boundary of the domain, we get
\begin{eqnarray}
\label{nr1b}
PV=Nk_BT-\frac{1}{2}{\cal V}-\frac{1}{2}\Omega L.
\end{eqnarray}
This is the equivalent of the usual virial theorem for material particles \cite{hansen}. However, for point vortices, it cannot be directly derived from the equations of motion (see the discussion at the end of Section \ref{exv}).  Using the expression (\ref{exv4}) of the virial (valid when vortex images are ignored) we obtain
\begin{eqnarray}
\label{nr2}
P V=Nk_B (T-T_c)-\frac{1}{2}\Omega L,
\end{eqnarray}
where $T_c$ is the critical temperature
\begin{eqnarray}
\label{nr3}
k_{B}T_{c}=-\frac{1}{8\pi N}\sum_{i=1}^N\sum_{j\neq i}\gamma_i\gamma_j=-\frac{1}{8\pi N}\left (\Gamma^2-\sum_{i=1}^N \gamma_i^2\right )=-\frac{1}{8\pi N}\left (\Gamma^2-\Gamma_2\right ).
\end{eqnarray}
For a neutral system
\begin{eqnarray}
\label{nr3bis}
k_{B}T_{c}=\frac{\Gamma_2}{8\pi N}.
\end{eqnarray}
For a system consisting of $N_+$ vortices of circulation $+\gamma$ and $N_-$ vortices of circulation $-\gamma$, we get
\begin{eqnarray}
\label{nr4}
k_{B}T_{c}=\frac{\gamma^2}{8\pi}\left\lbrack 1-\frac{(N_+-N_-)^2}{N}\right\rbrack.
\end{eqnarray}
In particular, for a neutral system with $N_+=N/2$ vortices $+\gamma$ and $N_-=N/2$ vortices $-\gamma$ , the critical temperature is
\begin{eqnarray}
\label{vc6b}
k_{B}T_{c}=\frac{\gamma^{2}}{8\pi}.
\end{eqnarray}
On the other hand, for a single species system with $N$ vortices $\gamma$, we get
\begin{eqnarray}
\label{vc6c}
k_{B}T_{c}=-(N-1)\frac{\gamma^{2}}{8\pi}.
\end{eqnarray}
If the domain has not the rotational symmetry, then $\Omega=0$. In that case, the virial theorem reduces to
\begin{eqnarray}
\label{nr5}
P V=Nk_B (T-T_c).
\end{eqnarray}
Therefore, the virial theorem determines the exact equation of state of a 2D system with logarithmic interactions. Since $PV/Nk_BT\ge 0$, we conclude that a statistical equilibrium state can possibly exist only if $\beta/\beta_c=T_c/T\le 1$. In particular, for a single species system for which $k_B T_c=-(N-1)\gamma^2/8\pi$ is negative, we must have $\beta\ge \beta_c$ ($T\ge 0$ or $T\le T_c$). On the other hand, for a neutral system for which $k_B T_c=\Gamma_2/8\pi N$ is positive, we must have $\beta\le \beta_c$ ($T\le 0$ or $T\ge T_c$). In the mean field approximation, we recover the results of Sec. \ref{sec_vmf}.

The preceding results are valid in the canonical ensemble. In the microcanonical ensemble, starting from Eq. (\ref{ym3}) and using the same procedure as before, we obtain the virial identity
\begin{eqnarray}
\label{mnr1b}
PV=(N-\alpha)k_BT-\frac{1}{2}{\cal V}-\frac{1}{2}k_B T\frac{\partial {\cal V}}{\partial E}-\frac{1}{2}\Omega L,
\end{eqnarray}
with $\alpha=1$ in a disk \footnote{In the case of an infinite domain treated in Secs. \ref{vmu} and \ref{sec_vvg}, we have seen that the conservation of angular momentum and the conservation of the center of vorticity both bring a term $k_B T$ in the virial theorem, leading to $\alpha=2$. In a disk, only the angular momentum is conserved so that $\alpha=1$. In a domain without particular symmetry, $\alpha=0$.}  and $\alpha=0$ in a domain that does not have the rotational symmetry ($\Omega L=0$).  Using the expression (\ref{exv4}) of the virial (valid when vortex images are ignored), we obtain
\begin{eqnarray}
\label{mnr2}
P V=Nk_B (T-T_c)-\frac{1}{2}\Omega L-\alpha k_B T,
\end{eqnarray}
where $T_c$ is the critical temperature (\ref{nr3}).

\subsection{From the partition function}
\label{sec_apf}

In the canonical ensemble, the pressure is defined  by
\begin{equation}
\label{apf1}
P=-\frac{\partial F}{\partial V}=\frac{1}{\beta}\frac{\partial \ln Z}{\partial V},
\end{equation}
where
\begin{eqnarray}
\label{apf2}
Z(\beta,V)=\int e^{\frac{\beta}{2\pi}\sum_{i<j}\gamma_{i}\gamma_{j}\ln |{\bf r}_{i}-{\bf r}_{j}|}\,  d{\bf r}_{1}...d{\bf r}_{N},
\end{eqnarray}
is the partition function (ignoring the contribution of vortex images). We have assumed that the system has not the rotational symmetry so that $\Omega=0$. If we make the change of variables ${\bf x}_{i}={\bf r}_{i}/\sqrt{V}$, the partition function becomes
\begin{eqnarray}
\label{apf3}
Z(\beta,V)=V^{N} e^{\frac{\beta}{4\pi}\ln V \sum_{i<j}\gamma_{i}\gamma_{j}}\int  e^{\frac{\beta}{2\pi}\sum_{i<j}\gamma_{i}\gamma_{j}\ln |{\bf x}_{i}-{\bf x}_{j}|}\, d{\bf x}_{1}...d{\bf x}_{N},
\end{eqnarray}
where the integral is independent on $V$. From this expression, we get
\begin{eqnarray}
\label{apf4}
\frac{\partial \ln Z}{\partial V}=\frac{N}{V}+\frac{\beta}{4\pi V}\sum_{i<j}\gamma_{i}\gamma_{j}.
\end{eqnarray}
Substituting this relation in Eq. (\ref{apf1}), we obtain the equation of state
\begin{eqnarray}
\label{apf5}
PV=Nk_{B}(T-T_{c}),
\end{eqnarray}
with the critical temperature (\ref{nr3}). This result extends to the multi-species case the calculations of Salzberg \& Prager \cite{spr,salzberg}.

For an arbitrary potential of interaction $u$, the partition function is given by
\begin{eqnarray}
\label{apf3new1}
Z(\beta,V)=\int  e^{-\beta H}\, d{\bf r}_{1}...d{\bf r}_{N}=\int  e^{-\beta\sum_{i<j}\gamma_{i}\gamma_{j} u(|{\bf r}_{i}-{\bf r}_{j}|)}\, d{\bf r}_{1}...d{\bf r}_{N}.
\end{eqnarray}
Making the change of variables ${\bf x}_{i}={\bf r}_{i}/\sqrt{V}$, we get
\begin{eqnarray}
\label{apf3new2}
Z(\beta,V)=V^{N}\int  e^{-\beta\sum_{i<j}\gamma_{i}\gamma_{j} u(\sqrt{V}|{\bf x}_{i}-{\bf x}_{j}|)}\, d{\bf x}_{1}...d{\bf x}_{N}.
\end{eqnarray}
Taking the derivative of this expression with respect to $V$, using Eq. (\ref{apf1}), and restoring the original coordinates, we obtain
\begin{eqnarray}
\label{apf3new3}
PV=Nk_B T-\frac{1}{2} \int \, d{\bf r}_{1}...d{\bf r}_{N} \frac{1}{Z(\beta,V)} e^{-\beta H} \sum_{i<j}\gamma_{i}\gamma_{j} |{\bf r}_{i}-{\bf r}_{j}| u'(|{\bf r}_{i}-{\bf r}_{j}|).
\end{eqnarray}
In the last term, we recognize the canonical average of the virial (\ref{exv2tris}). Therefore, Eq. (\ref{apf3new3}) can be rewritten
\begin{eqnarray}
\label{apf3new4}
PV=Nk_B T-\frac{1}{2}{\cal V}.
\end{eqnarray}
Using the expression (\ref{exv4}) of the virial (valid when vortex images are ignored) we recover Eq. (\ref{apf5}). We note that Eq. (\ref{apf3new4}) coincides with the virial theorem (\ref{nr1b}) obtained from the YBG equation. Therefore, we conclude that the pressure $P$ defined by Eq. (\ref{nr1}) with the isothermal equation of state (\ref{vc1}) can be identified with the thermodynamical pressure. This was not obvious {\it a priori}. Therefore, even if the approach developed in this section is not rigorous (since it ignores vortex images), it shows a ``consistency'' between the different definitions of pressure that we have given in the paper.

\subsection{From the density of states}
\label{ads}

In the microcanonical ensemble, the pressure and the temperature are defined by
\begin{equation}
\label{ads1}
P=T\frac{\partial S}{\partial V}, \qquad \frac{1}{T}=\frac{\partial S}{\partial E},
\end{equation}
where $S(E,V)=k_{B}\ln g(E,V)$ is the entropy and $g(E,V)$  the density of states
\begin{eqnarray}
\label{ads2}
g(E,V)=\int \delta \biggl (E+\frac{1}{2\pi}\sum_{i<j}\gamma_{i}\gamma_{j}\ln |{\bf r}_{i}-{\bf r}_{j}|\biggr ) \, d{\bf r}_{1}...d{\bf r}_{N}.
\end{eqnarray}
We have assumed that the system has not the rotational symmetry so there is no constraint on the angular momentum. With the change of variables ${\bf x}_{i}={\bf r}_{i}/\sqrt{V}$, the density of states can be rewritten
\begin{eqnarray}
\label{ads3}
g(E,V)=V^{N}\int \delta \biggl (E'+\frac{1}{2\pi}\sum_{i<j}\gamma_{i}\gamma_{j}\ln |{\bf x}_{i}-{\bf x}_{j}|\biggr )\, d{\bf x}_{1}...d{\bf x}_{N},
\end{eqnarray}
where
\begin{eqnarray}
\label{ads4}
E'=E+\frac{1}{4\pi}\ln V \sum_{i<j}\gamma_{i}\gamma_{j}.
\end{eqnarray}
Therefore, the density of states is of the form
\begin{eqnarray}
\label{ads5}
g(E,V)=V^{N}g(E',1),
\end{eqnarray}
and the entropy  is of the form
\begin{eqnarray}
\label{ads6}
S(E,V)=N k_{B}\ln V+S(E',1).
\end{eqnarray}
From this expression, we obtain
\begin{eqnarray}
\label{ads7}
\frac{\partial S}{\partial V}=Nk_B \frac{1}{V}+\frac{\partial S}{\partial E}(E',1)\frac{1}{4\pi V}\sum_{i<j}\gamma_{i}\gamma_{j},
\end{eqnarray}
and
\begin{eqnarray}
\label{ads8}
\frac{\partial S}{\partial E}=\frac{\partial S}{\partial E}(E',1).
\end{eqnarray}
Using Eq. (\ref{ads1}) we get
\begin{eqnarray}
\label{ads9}
PV=Nk_{B}(T-T_{c}),
\end{eqnarray}
with the critical temperature (\ref{nr3}). This result extends to the multi-species case the calculations of  Edwards \& Taylor \cite{et}.

For an arbitrary potential of interaction $u$, the density of states is given by
\begin{eqnarray}
\label{apf3new5}
g(E,V)=\int \delta (E-H) \, d{\bf r}_{1}...d{\bf r}_{N}=\int \delta \biggl (E-\sum_{i<j}\gamma_{i}\gamma_{j} u(|{\bf r}_{i}-{\bf r}_{j}|)\biggr ) \, d{\bf r}_{1}...d{\bf r}_{N}.
\end{eqnarray}
Making the change of variables ${\bf x}_{i}={\bf r}_{i}/\sqrt{V}$, we get
\begin{eqnarray}
\label{apf3new6}
g(E,V)=V^N\int \delta \biggl (E-\sum_{i<j}\gamma_{i}\gamma_{j} u(\sqrt{V}|{\bf r}_{i}-{\bf r}_{j}|)\biggr ) \, d{\bf x}_{1}...d{\bf x}_{N}.
\end{eqnarray}
Taking the derivative of this expression with respect to $V$, using Eq. (\ref{ads1}), and restoring the original coordinates, we obtain
\begin{eqnarray}
\label{apf3new7}
\frac{P V}{k_B T}=N-\frac{1}{2}\frac{1}{g(E,V)}\frac{\partial}{\partial E} g(E,V)\int \, d{\bf r}_{1}...d{\bf r}_{N} \frac{1}{g(E,V)}\delta(E-H) \sum_{i<j}\gamma_{i}\gamma_{j} |{\bf r}_{i}-{\bf r}_{j}| u'(|{\bf r}_{i}-{\bf r}_{j}|).
\end{eqnarray}
In the last term, we recognize the microcanonical average of the virial (\ref{exv2tris}). Therefore, Eq. (\ref{apf3new7}) can be rewritten
\begin{eqnarray}
\label{apf3new8}
\frac{P V}{k_B T}=N-\frac{1}{2}\frac{1}{g(E,V)}\frac{\partial}{\partial E} (g(E,V){\cal V}).
\end{eqnarray}
Expanding the derivative, and using Eq. (\ref{ads1}), we obtain
\begin{eqnarray}
\label{apf3new9}
P V=Nk_B T-\frac{1}{2}{\cal V}-\frac{1}{2}k_B T\frac{\partial {\cal V}}{\partial E}.
\end{eqnarray}
Using the expression (\ref{exv4}) of the virial (valid when vortex images are ignored) we recover Eq. (\ref{ads9}). We note that Eq. (\ref{apf3new9}) coincides with the virial theorem (\ref{mnr1b}) obtained from the YBG equation. This allows us to identify expression (\ref{nr1}) with the thermodynamical pressure. 

Finally, we note that the equations of state (\ref{apf5}) and (\ref{ads9}) have the same form in the microcanonical and canonical ensembles (recall, however, that in the microcanonical ensemble $T=T(E)$ is a function of the energy while in the canonical ensemble $T$ is prescribed). This equivalence is not expected to be always true because, for an arbitrary potential of interaction $u$, the expressions (\ref{apf3new4}) and (\ref{apf3new9}) differ by a factor $-({1}/{2})k_B T {\partial {\cal V}}/{\partial E}$ that may not vanish. However, in the limit $N\rightarrow +\infty$, this term is negligible and the equations of state always coincide in the microcanonical and canonical ensembles.

\section{The case of a neutral spatially homogeneous system}
\label{sec_yy}

We consider here a neutral system made of $N/2$ point vortices of
circulation $+\gamma$ and $N/2$ point vortices of circulation
$-\gamma$. We assume that the system is spatially homogeneous. The
density of each species is therefore $n_+=n_{-}=n/2$ where $n=N/V$ is
the total density.   As shown in Appendix \ref{sec_nonrig}, the virial theorem leads to the equation of state
\begin{eqnarray}
\label{yy2}
PV=Nk_{B}(T-T_{c}),
\end{eqnarray}
with
\begin{eqnarray}
\label{yy3}
k_{B}T_{c}=\frac{\gamma^2}{8\pi}.
\end{eqnarray}

Let us first consider positive temperatures. In that case, the system
is always spatially homogeneous. The exact equation of state
(\ref{yy2})-(\ref{yy3}) was originally derived by Salzberg \& Prager
\cite{spr,salzberg} and May
\cite{may} for a spatially homogeneous 2D neutral plasma.
According to Eq. (\ref{yy2}), the pressure is negative for $T<T_c$,
which is not possible at positive temperatures. Actually, it can be
shown that the partition function is convergent only for
$T>T_*=\gamma^2/(4\pi)=2T_c$
\cite{knorr,haugehemmer,deutschlavaud1,gonzalezhemmer,deutschlavaud2,exact}
(see Appendix \ref{er}) so the equation of state (\ref{yy2}) is not valid
for $T<T_*$. For $T<T_*$, there is no equilibrium state and the system
``collapses''. This leads to the formation of $N/2$ non-interacting
singular pairs $(+,-)$ of opposite charge, corresponding atoms
$(+e,-e)$ in plasma physics or to dipoles $(+\gamma,-\gamma)$ in 2D
hydrodynamics. For $T=T_*$, we have $PV=(1/2)Nk_B T_*$. This is the
equation of state of an ideal gas of $N/2$ non-interacting singular
pairs $(+,-)$.

We write the two-body distribution function as $P_2^{(ab)}(|{\bf
r}-{\bf r}'|)=P_0^2+{P_2'}^{(ab)}(|{\bf r}-{\bf r}'|)$ where
$P_0=1/V$ is the spatially homogeneous one-body distribution and ${P_2'}^{(ab)}(|{\bf r}-{\bf r}'|)$ is the two-body correlation function. Due to the symmetry of charges $+\gamma$ and $-\gamma$,  we have
${P_2'}^{(++)}={P_2'}^{(--)}=-{P_2'}^{(+-)}=-{P_2'}^{(-+)}=P_0^2
h(|{\bf r}-{\bf r}'|)$. Using the second equation of the YBG hierarchy
and implementing a Debye-H\"uckel approximation which consists in
neglecting three-body correlations, it can be shown  that the two-body correlation function $h_{DH}(|{\bf r}-{\bf r}'|)$ is the solution
of the differential equation \cite{vahala,monthouches,pl,bv}:
\begin{eqnarray}
\label{yy4}
\Delta h_{DH}-k_D^2 h_{DH}=\beta\gamma^2 \delta({\bf x}),
\end{eqnarray}
where $k_D=(\beta n\gamma^2)^{1/2}$ is the Debye wavenumber. The Debye-H\"uckel approximation is valid at sufficiently high temperatures $T$ (small $\beta$). In particular, it does not signal the collapse of pairs $(+,-)$ at $T<T_*$.  The solution of Eq. (\ref{yy4}) is
\begin{eqnarray}
\label{yy5}
n (2\pi)^2 \hat{h}_{DH}(k)=\frac{-k_D^2}{k^2+k_D^2},\qquad h_{DH}(x)=-\frac{\beta \gamma^2}{2\pi}K_{0}(k_D x).
\end{eqnarray}
This result can also be obtained from a diagrammatic expansion  \cite{deutschlavaud3,deutschlavaud2} in terms of the plasma parameter $\epsilon=\beta\gamma^2/2\pi\ll 1$. A more precise expression of the correlation function (obtained from a straightforward renormalization of the short-range behavior of $h$) is given by $h(x)=1-e^{\epsilon K_0(k_D x)}$ \cite{deutschlavaud3,deutschlavaud2}. According to Eq. (\ref{yy5}), a vortex of a given sign is surrounded by vortices of opposite sign which screen its interaction on a distance of the order of the Debye length $\sim k_D^{-1}$.

Let us now consider the case of negative temperatures. In that case,
the system is spatially homogeneous above an inverse temperature
$\beta_0<0$ (see the Remark at the end of Sect. \ref{sec_bpy}).  The
exact equation of state (\ref{yy2})-(\ref{yy3}) is still valid
\cite{et} and we note that the pressure is negative (which is allowed
at negative temperatures). The Debye-H\"uckel approximation can also
be extended to negative temperatures in the range $\beta_0<\beta<0$. The two-body correlation function $h_{DH}(|{\bf r}-{\bf r}'|)$ is the solution
of the differential equation \cite{bv}:
\begin{eqnarray}
\label{yy6}
\Delta h_{DH}+k_J^2 h_{DH}=\beta\gamma^2 \delta({\bf x}),
\end{eqnarray}
where $k_J=(-\beta n\gamma^2)^{1/2}$ is the equivalent of the Jeans wavenumber in astrophysics \cite{jeans}.  The solution of this equation is
\begin{eqnarray}
\label{yy7}
n (2\pi)^2 \hat{h}_{DH}(k)=\frac{k_J^2}{k^2-k_J^2},\qquad h_{DH}(x)=\frac{\beta \gamma^2}{4}Y_{0}(k_J x).
\end{eqnarray}
The oscillatory behavior of the correlation function $h_{DH}(x)$, or
the fact that the denominator of $\hat{h}_{DH}(x)$ is negative for
$k<k_J$, is related to the clustering of point vortices at negative
temperatures \cite{bv}. This is consistent with the results of Edwards
\& Taylor \cite{et} obtained in a different manner. For smaller values
of the inverse temperature $\beta$, the system is spatially
inhomogeneous (forming a large-scale dipole) and we must use the mean
field results of Sect. \ref{sec_mfa} that are exact at the
thermodynamic limit. For $\beta<\beta_*=-16\pi/N\gamma^2$, there is no
equilibrium state anymore and the system forms a singular macroscopic
dipole $(\frac{N}{2}+,\frac{N}{2}-)$ (see Appendix \ref{er}).

We now consider the virial equation \cite{hansen}:
\begin{eqnarray}
\label{yy8}
PV=Nk_BT-\frac{1}{4}n^2\gamma^2 V\int_{0}^{+\infty} g(x) x u'(x) 2\pi x \, dx,
\end{eqnarray}
where $u(x)$ is the potential of interaction (\ref{pv3}). This equation can be directly obtained from Eqs. (\ref{nr1b}) and (\ref{exv8bis}), and it leads to Eq. (\ref{yy2}).  If we replace the correlation function $g(x)=1+h(x)$ by the Debye-H\"uckel expression $h_{DH}(x)$ given by Eqs. (\ref{yy5}) and (\ref{yy7}), and  carry out the integration, we obtain Eq. (\ref{yy2}). Therefore, the virial equation calculated with the Debye-H\"uckel correlation function returns the {\it exact} equation of state of a 2D plasma or a 2D point vortex gas. However, this is essentially a coincidence since Eq. (\ref{yy8}) is generally valid for the correlation function $g(x)$, not for $h_{DH}(x)$.

\section{Existence of statistical equilibrium states}
\label{er}

We first consider a system of $N$ point vortices of equal circulation $\gamma$ in a bounded domain. At positive temperatures $\beta>0$, like-sign vortices ``repel'' each other and accumulate on the boundary of the domain. As the inverse temperature increases (or as the energy decreases), the vortices are more an more concentrated on the boundary.  This regime of low energies corresponds to a repulsive interaction between vortices, like between electric charges of the same sign in electrolytes \cite{dh}.  At negative temperatures $\beta<0$, like-sign vortices ``attract'' each other and form clusters. As the inverse temperature decreases (or as the energy increases), the cluster is more and more condensed. This regime of high energies corresponds to an attractive interaction between vortices, like between stars in a galaxy \cite{paddy,ijmpb}. It can be shown \cite{caglioti,k93,exact} that statistical equilibrium states exist (the partition function is finite) if, and only, if
\begin{eqnarray}
\label{er1}
\beta>\beta_{*}^{(-)}=-\frac{8\pi}{N\gamma^2}.
\end{eqnarray}
When $\beta\rightarrow +\infty$, all the vortices concentrate on the boundary. When $\beta\rightarrow  \beta_{*}^{(-)}$, all the vortices collapse and form a Dirac peak $\omega({\bf r})=\Gamma\delta({\bf r})$. In that case, $E\rightarrow +\infty$. In an unbounded domain, the picture is slightly different (see Sec. \ref{sec_cons} for more detailed results in the mean field limit). At positive temperatures $\beta>0$, like-sign vortices ``repel'' each other but the spreading of the vortices is prevented by the conservation of angular momentum in MCE or by the confining potential $\psi_{conf}=\frac{\Omega}{2}r^2$ in CE. When $\beta\rightarrow +\infty$, the vorticity profile is a step function (in the mean field limit).  At negative temperatures $\beta_{*}^{(-)}<\beta<0$, like-sign vortices ``attract'' each other and form clusters. As the inverse temperature decreases, the cluster is more and more condensed. When $\beta\rightarrow  \beta_{*}^{(-)}$, the vorticity profile is a Dirac peak in CE and a Dirac peak surrounded by a residual vorticity profile with a Gaussian tail  (in the mean field limit) which allows to satisfy the conservation of angular momentum in MCE. In that case, $E\rightarrow +\infty$.

We now consider a neutral system made of $N/2$ point vortices of
circulation $+\gamma$ and $N/2$ point vortices of circulation
$-\gamma$ in a bounded domain.  At positive temperatures $\beta>0$,
the interaction between like-sign vortices is ``repulsive'' and the
interaction between opposite-sign vortices is ``attractive''. This is
similar to the case of electric charges in electrolytes \cite{dh}.  At
high temperatures (low $\beta$), each point vortex is surrounded by a
cloud of opposite-sign vortices which screens the interaction
\cite{et,bv}. This is similar to the Debye shielding in plasma physics
\cite{dh}. In that case, the system is spatially homogeneous and fully
``ionized''. At low temperatures (high $\beta$), the vortices tend to
form pairs of microscopic dipoles $(+,-)$ similar to ``atoms''
$(+e,-e)$ in plasma physics. The system consists in a spatially
homogeneous gas of non-interacting microscopic dipoles. This
corresponds to the regime of low energies. At negative temperatures
$\beta<0$, the interaction between like-sign vortices is
``attractive'' and the interaction between opposite-sign vortices is
``repulsive''. At small $\beta$, the system is spatially homogeneous
in average but the vortices tend to cluster. This corresponds to a
form of anti-shielding in which each point vortex is surrounded by a
cloud of similar vortices \cite{et,bv}. This is analogous to the Jeans
gravitational clustering in astrophysics \cite{jeans}. At lower
$\beta$, the vortices form two (or more) macroscopic clusters
(dipoles, tripoles,...) made of vortices of the same sign
\cite{onsager,jm}.  In that case, the system is spatially
inhomogeneous and the structures can be described within the mean
field theory which is exact. This corresponds to the regime of high
energies. It can be shown
\cite{km,exact} that statistical equilibrium states exist (the partition function is finite)  if, and only, if
\begin{eqnarray}
\label{er2}
\beta_{*}^{(-)}=-\frac{16\pi}{N\gamma^2}<\beta<\beta_{*}^{(+)}=\frac{4\pi}{\gamma^2}.
\end{eqnarray}
When $\beta\rightarrow \beta_{*}^{(-)}$, the system forms a singular
macroscopic dipole $(\frac{N}{2}+,\frac{N}{2}-)$ with one Dirac peak
$+\frac{N}{2}\gamma$ containing all the positive vortices and one
Dirac peak $-\frac{N}{2}\gamma$ containing all the negative
vortices. In that case, $E\rightarrow +\infty$. The critical
temperature $\beta_{*}^{(-)}$ can be understood by considering the
statistical mechanics of a single species system of $N/2$ vortices
(since each peak contains $N/2$ vortices). It can be therefore deduced
from Eq. (\ref{er1}) by making the substitution $N\rightarrow
N/2$. When $\beta\rightarrow \beta_{*}^{(+)}$, the system forms $N/2$
singular microscopic dipoles $(+,-)$ where the vortices of each pair
have fallen on each other. This leads to $N/2$ Dirac peaks made of two
vortices of opposite circulation. In that case, $E\rightarrow
-\infty$. The critical temperature $\beta_{*}^{(+)}$ can be understood
by considering the statistical mechanics of only two vortices of
opposite sign. It can therefore be deduced from Eq. (\ref{er1}) by
taking $N=2$ (since only two vortices are involved in the dipole) and
reversing the sign (since the attraction of like-sign vortices at
negative temperatures corresponds to the attraction of opposite-sign
vortices at positive temperatures).

For a multi-species system of point vortices in a bounded domain, the conditions of existence of the statistical equilibrium state, and the corresponding expressions of the critical temperature, are more difficult to obtain. Some interesting mathematical results have been obtained by Ohtsuka {\it et al.}  \cite{ors} in the mean field approximation. Explicit expressions of the critical temperature have also been obtained by Sopik {\it et al.}  \cite{sopik} when the system consists of two types of like-sign vortices (this study is performed in the context of self-gravitating systems, but it can be readily applied to point vortices at negative temperatures).






\end{document}